\documentclass[onefignum,onetabnum]{siamonline190516}

\usepackage{braket,amsfonts}

\usepackage{array}

\usepackage[caption=false]{subfig}
\usepackage{pgfplots}

\newsiamthm{claim}{Claim}
\newsiamthm{remark}{Remark}
\newsiamthm{hypothesis}{Hypothesis}
\crefname{hypothesis}{Hypothesis}{Hypotheses}
\usepackage{booktabs}

\usepackage{algorithm}
\usepackage[noend]{algpseudocode}
\usepackage{etoolbox}

\usepackage{graphicx,epstopdf}

\Crefname{ALC@unique}{Line}{Lines}

\usepackage{amsopn}

\usepackage{xspace}
\usepackage{bold-extra}
\usepackage[most]{tcolorbox}

\colorlet{texcscolor}{blue!50!black}
\colorlet{texemcolor}{red!70!black}
\colorlet{texpreamble}{red!70!black}
\colorlet{codebackground}{black!25!white!25}

\lstdefinestyle{siamlatex}{%
  style=tcblatex,
  texcsstyle=*\color{texcscolor},
  texcsstyle=[2]\color{texemcolor},
  keywordstyle=[2]\color{texemcolor},
  moretexcs={cref,Cref,maketitle,mathcal,text,headers,email,url},
}

\makeatletter
\def\BState{\State\hskip-\ALG@thistlm}

\newcommand{\distas}[1]{\mathbin{\overset{#1}{\kern\z@\sim}}}%
\newcommand{\bm}[1]{\mathbf{#1}}
\newsavebox{\mybox}\newsavebox{\mysim}
\newcommand{\distras}[1]{%
  \savebox{\mybox}{\hbox{\kern3pt$\scriptstyle#1$\kern3pt}}%
  \savebox{\mysim}{\hbox{$\sim$}}%
  \mathbin{\overset{#1}{\kern\z@\resizebox{\wd\mybox}{\ht\mysim}{$\sim$}}}%
}

\newcommand{\be}{\begin{equation}}
\newcommand{\ee}{\end{equation}}
\newcommand{\bi}{\begin{itemize}}
\newcommand{\ei}{\end{itemize}}
\newcommand{\ben}{\begin{enumerate}}
\newcommand{\een}{\end{enumerate}}
\newcommand{\stb}{\State $\bullet$ \;}

\newcolumntype{K}[1]{>{\centering\arraybackslash}p{#1}}
\allowdisplaybreaks
\DeclareMathOperator*{\argmin}{\arg\!\min}
\makeatother

\tcbset{%
  colframe=black!75!white!75,
  coltitle=white,
  colback=codebackground, 
  colbacklower=white, 
  fonttitle=\bfseries,
  arc=0pt,outer arc=0pt,
  top=1pt,bottom=1pt,left=1mm,right=1mm,middle=1mm,boxsep=1mm,
  leftrule=0.3mm,rightrule=0.3mm,toprule=0.3mm,bottomrule=0.3mm,
  listing options={style=siamlatex}
}

\newtcblisting[use counter=example]{example}[2][]{%
  title={Example~\thetcbcounter: #2},#1}

\newtcbinputlisting[use counter=example]{\examplefile}[3][]{%
  title={Example~\thetcbcounter: #2},listing file={#3},#1}

\DeclareTotalTCBox{\code}{ v O{} }
{ 
  fontupper=\ttfamily\color{black},
  nobeforeafter,
  tcbox raise base,
  colback=codebackground,colframe=white,
  top=0pt,bottom=0pt,left=0mm,right=0mm,
  leftrule=0pt,rightrule=0pt,toprule=0mm,bottomrule=0mm,
  boxsep=0.5mm,
  #2}{#1}

\patchcmd\newpage{\vfil}{}{}{}
\begin{tcbverbatimwrite}{tmp_\jobname_header.tex}
\title{APIK: Active Physics-Informed Kriging Model with Partial Differential Equations}

\author{Jialei Chen\thanks{The H. Milton Stewart School of Industrial and Systems Engineering, Georgia Institute of Technology, Atlanta,
GA 30332 (\email{jialei.chen@gatech.edu}, \email{zhchen@gatech.edu}, \email{chuck.zhang@gatech.edu}, \email{jeff.wu@isye.gatech.edu}).}
\and Zhehui Chen\footnotemark[1]
\and Chuck Zhang\footnotemark[1]~\thanks{Georgia Tech Manufacturing Institute.}
\and C. F. Jeff Wu\footnotemark[1]}

\headers{Kriging with Partial Differential Equations}{Jialei Chen, Zhehui Chen, Chuck Zhang and C. F. Jeff Wu}
\end{tcbverbatimwrite}
\input{tmp_\jobname_header.tex}

\ifpdf
\hypersetup{ pdftitle={APIK} }
\fi

\begin{document}
\maketitle

\begin{tcbverbatimwrite}{tmp_\jobname_abstract.tex}
\begin{abstract}
Kriging (or Gaussian process regression) is a popular machine learning method for its flexibility and closed-form prediction expressions.
However, one of the key challenges in applying kriging to engineering systems is that the available measurement data is scarce due to the measurement limitations and high sensing costs. 
On the other hand, physical knowledge of the engineering system is often available and represented in the form of partial differential equations (PDEs).
We present in this work a PDE Informed Kriging model (PIK), which introduces PDE information via a set of PDE points and conducts posterior prediction similar to the standard kriging method. 
The proposed PIK model can incorporate physical knowledge from both linear and nonlinear PDEs.
To further improve learning performance, we propose an Active PIK framework (APIK) that designs PDE points to leverage the PDE information based on the PIK model and measurement data.
The selected PDE points not only explore the whole input space but also exploit the locations where the PDE information is critical in reducing predictive uncertainty. Finally, an expectation-maximization algorithm is developed for parameter estimation. We demonstrate the effectiveness of APIK in two synthetic examples, a shock wave case study, and a laser heating case study.  

\end{abstract}


\end{tcbverbatimwrite}
\input{tmp_\jobname_abstract.tex}

\section{Introduction}
\label{sec:intro}

Recent advances in machine learning have attracted increasing attention from the engineering community. 
Among these learning methods, \textit{kriging} (or Gaussian process regression) \cite{williams2006gaussian,santner2018design} is particularly popular for its flexible structure and the fact that both the prediction and its uncertainty quantification enjoy simple and closed-form expressions. 
Therefore, kriging has been adopted in a variety of applications in different domains, including geostatistics \cite{matheron1963principles,diggle1998model}, financial engineering \cite{sottinen2001fractional,hernandez2013gaussian}, and computer experiment \cite{sacks1989design,currin1991bayesian}.
Despite the popularity, one of the key challenges in real-world engineering applications is that the available amount of data is often extremely limited due to the constraints of measurement devices and human labor.
For example, in a laser heating process in wafer manufacturing \cite{chen2019adaptive,chen2019hierarchical}, the goal is to understand the temperature profile of a rotating wafer heated by a laser beam (more details in \cref{sec:HeatTrans}).
The temperature measurements rely on a handheld temperature gun. 
Only a handful of readings during the heating process is available since (i) the measurement procedure is labor-intensive, and (ii) the response time of the temperature gun is around one second. To address this, we present in this work a new kriging modeling framework, which can improve predictive performance with limited data.

\begin{figure}
\centering
\includegraphics[width=0.7\textwidth]{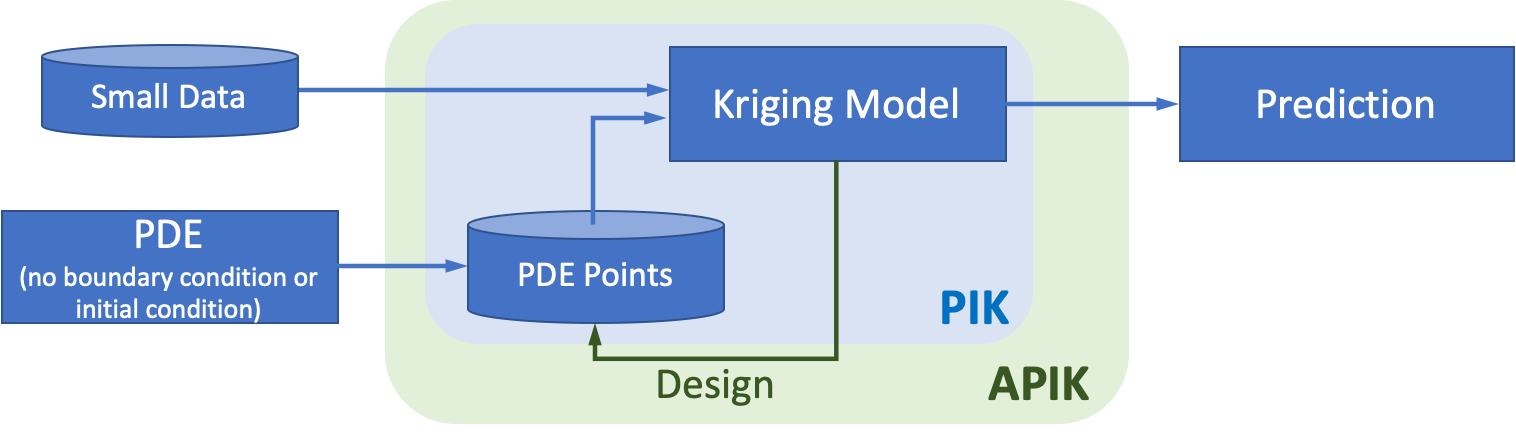}
\caption{\label{fig:ill} An illustration of the proposed PIK and APIK methods. The PIK method can incorporate the PDE information in the kriging framework, and the APIK model can actively select PDE points to leverage the PDE information according to the measurement data.}
\end{figure}

One way to improve learning performance (we adopt) is to incorporate \textit{physical knowledge}, which can also be viewed as enlarging the existing data set \cite{willard2020integrating,han2020integrating}.  
Recently, researchers have also proposed several physics-informed kriging methods to incorporate physical knowledge, 
such as bound constraints \cite{da2012gaussian}, invariance equations \cite{chen2020function}, monotonicity information \cite{riihimaki2010gaussian}, and output's correlation structure \cite{mak2018efficient}. However, the physical knowledge considered in those works only exists on a \textit{case-by-case} basis.
In this work, we develop a kriging framework with partial differential equation (PDE) shown in \cref{fig:ill}, which is a rigorous representation of the first principle and exists for almost \textit{all} engineering systems. 
Take the previous wafer manufacturing application as an example. Besides the limited data, we know that the temperature profile of the wafer is governed by the Fourier equation \cite{eslami2013theory}, a PDE describing macroscopic transport properties of energy.
Our goal is to incorporate the Fourier equation with measurement data, thereby better predicting the temperature profile.

There are several related works in the kriging literature. Wang and Berger \cite{wang2016estimating} developed a kriging models to incorporate derivative constraints, {\it i.e.}, the simplest PDE information, with follow-up works in \cite{ulaganathan2016performance,han2017weighted,he2018instability}. 
Wheeler {\it et al.}~\cite{wheeler2014mechanistic} proposed a hierarchical Gaussian process model that favors curves consistent with the considered linear PDEs.
After that, Jidling {\it et al.}~\cite{jidling2017linearly} developed a kriging model with modification of the correlation function to account for linear PDEs.
Raissi {\it et al.}~\cite{raissi2017machine} proposed a kriging-based method to evaluate the unknown coefficients in linear PDEs, with further development in \cite{gulian2019machine}. 
Besides, Raissi {\it et al.}~\cite{raissi2018numerical} presented a kriging model for the evolution of dynamic systems with a correlation structure inherited from PDEs.
There are also works on incorporating boundary condition constraints \cite{solin2019know,ding2019bdrygp}.
However, to the best of our knowledge, there is no systematic modeling framework for incorporating \textit{nonlinear} PDEs into kriging methods.

In this work, we propose a new kriging model to fuse general PDEs and measurement data. We develop first a PDE Informed Kriging model (PIK) to utilize a set of pseudo points, called  \textit{PDE points}, to incorporate physical knowledge from linear PDEs and nonlinear PDEs. Specifically, for linear PDEs, we extend the learning method of incorporating gradient information in \cite{wang2016estimating}. For nonlinear PDEs, we introduce a set of latent variables, which converts the nonlinear PDEs to linear PDEs by the conditional process. 
To further improve the learning performance, we propose an Active PIK framework (APIK) that leverages the PDE information by actively designing the PDE points based on the PIK model and measurement points. The selected PDE points not only explore the whole input space but also exploit the locations where the PDE information is the most important in reducing predictive uncertainty. We then develop an expectation-maximization method for APIK, which efficiently estimates model parameters and latent variables, as well as actively selects the locations for PDE points. 
Further discussions on PDE data size, computational simplification, and the connection to numerical PDE solvers are included. 
Finally, we demonstrate the improvements of the proposed APIK method in both synthetic examples and real case studies.

The remaining part of the article is organized as follows. In \cref{sec:PIK}, we present the PIK model with linear and nonlinear PDEs. In \cref{sec:APIK}, we develop the APIK model based on PIK. We discuss a parameter estimation algorithm in \cref{sec:learnAPIK}. PIK and APIK are then applied to four applications in \cref{sec:example}.
Finally, \cref{sec:summary} concludes this work.

\section{PIK model}
\label{sec:PIK}
In this section, we first review the kriging method. We then present the PIK model with linear PDEs. Finally, we extend the PIK model to nonlinear PDEs. 

\subsection{Kriging model} \label{sec:GPregression}
Let $\mathbf{x}_i \in \mathcal{X}\subset \mathbb{R}^d$ be a vector of $d$ input variables in a bounded space $ \mathcal{X}$, and let $y_i\in \mathbb{R}$ be the corresponding measurement of the physical quantity. We consider the following model for the physical quantity of interest:
\begin{align}
y_i=y(\bm{x}_i)+\epsilon_i, \quad i = 1,2, \cdots, n,
\label{eq:DataModel}
\end{align}
where $y(\bm{x}_i)$ is the underlying output physical quantity at input $\bm{x}_i$, $\epsilon_i$ is the corresponding measurement error, and $n$ is the size of measurement data. The experimental noise $\epsilon_i \sim \mathcal{N}(0,\sigma^2_\epsilon)$ is assumed to be i.i.d. normally distributed and independent to $y(\cdot)$. Kriging model further supposes that the input-output relationship $y(\cdot): \mathcal{X}\mapsto\mathbb{R}$ follows a Gaussian process 
\begin{align}
   y(\mathbf{x})\sim \mathcal{GP}(\mu(\mathbf{x}),\sigma^2 R_{\boldsymbol\theta} (\cdot,\cdot)), \quad \text{with} \quad  \mu(\mathbf{x}) = \mathbf{p}^\top(\mathbf{x})\boldsymbol\beta
    \label{eq:NGP_model}.
\end{align}
Here, $\mathbf{p}(\mathbf{x}) = [p_1(\mathbf{x}),\cdots, p_q(\mathbf{x})]^\top$ consists of $q$ basis functions for the mean function $\mu(\mathbf{x})$, $\boldsymbol\beta\in\mathbb{R}^q$ denotes the corresponding coefficients, $\sigma^2$ is the process variance, and $R_{\boldsymbol\theta}(\cdot,\cdot)$ is the correlation function with parameter $\boldsymbol\theta$.  
Denote the response vector of measurement data  $\mathbf{y}_{1:n} = [y_1, \cdots, y_n]^\top$. We have the following proposition for the posterior prediction. 
\begin{proposition} [Page 17 in \cite{williams2006gaussian}]
\label{le:NGP}
Suppose that the Gaussian process model \eqref{eq:DataModel}-\eqref{eq:NGP_model} holds with known model parameters $\{\boldsymbol\beta, \sigma^2, \boldsymbol\theta\}$.
Then the posterior prediction $y(\bm{x}_{\rm new})|\mathbf{y}_{1:n}$ at a new input location $\mathbf{x}_{\rm new}$ is normally distributed with mean and variance
\begin{align}
     \mathbb{E} \left[y(\mathbf{x}_{\rm new})|\mathbf{y}_{1:n}\right] &=  \mathbf{p}^\top(\mathbf{x}_{\rm new})\boldsymbol\beta + \mathbf{r}^\top(\mathbf{x}_{\rm new}) \mathbf{R}_{\boldsymbol\theta}^{-1}(\mathbf{y}_{1:n}-\mathbf{P}\boldsymbol\beta), \label{eq:NGP_E}\\
\text{\rm Var} \left[y(\mathbf{x}_{\rm new})|\mathbf{y}_{1:n}\right] &=  \sigma^2 (1- \mathbf{r}^\top(\mathbf{x}_{\rm new})\mathbf{R}_{\boldsymbol\theta}^{-1}\mathbf{r}(\mathbf{x}_{\rm new})). \label{eq:NGP_var}
\end{align}
where, $\mathbf{P}= [\mathbf{p}(\mathbf{x}_1),\cdots, \mathbf{p}(\mathbf{x}_n)]^\top$, $\mathbf{r}(\mathbf{x}_{\rm new}) = [R_{\boldsymbol\theta}(\mathbf{x}_{\rm new},\mathbf{x}_1), \cdots, R_{\boldsymbol\theta}(\mathbf{x}_{\rm new},\mathbf{x}_n)]^\top$, and $\mathbf{R}_{\boldsymbol\theta}={[R_{\boldsymbol\theta}(\mathbf{x}_i,\mathbf{x}_j)]_{i=1}^n}_{j=1}^n+\sigma^2_\epsilon/\sigma^2\bm{I}_n$ with $\bm{I}_n$ the $n\times n$ identity matrix.
\end{proposition}
\noindent 
Here, $\mathbf{P}$ is the $n \times q$ model matrix at $n$ measurement points, $\mathbf{r}(\mathbf{x}_{\rm new})$ is the correlation vector between the $n$ points and the predictive location, and $\mathbf{R}_{\boldsymbol\theta}$ is the corresponding  $n\times n$ correlation matrix. The conditional mean in \eqref{eq:NGP_E} can be used for prediction, and the conditional variance in \eqref{eq:NGP_var} can be used for constructing point-wise predictive intervals.

\subsection{PIK with linear PDEs} \label{sec:Linear}
As discussed, PDE information on the physical quantity $y(\bm{x})$ is available. Consider for now the PDEs are linear; Nonlinear PDEs will be discussed later in \cref{sec:nonlinearPDE}. We have the following definition for \textit{linear} PDEs.
\begin{definition} [Linear PDE]
\label{def:linearPDE}
PDE $\mathcal{F}_{\bm{x}}[y](\bm{x}) = b(\bm{x})$ with a non-homogeneous term $b(\bm{x})$ is called a linear PDE if $\mathcal{F}_{\bm{x}}[y](\bm{x})$ is a linear differential operator:
\begin{align}
  \mathcal{F}_{\bm{x}}[y](\bm{x}) :=\sum_{i=1}^L c_i(\bm{x})\nabla_{\boldsymbol\alpha_i} y(\bm{x}),
  \quad
  \text{with}  \quad \nabla_{\boldsymbol\alpha_i} y := \frac{\partial^{\alpha_{i,1}} }{\partial {x}_1^{\alpha_{i,1}}}\cdots  \frac{\partial^{\alpha_{i,d}} }{\partial {x}_d^{\alpha_{i,d}}} y, 
  \label{eq:LinearPDE}
\end{align}
 $\boldsymbol\alpha_i = \left[\alpha_{i,1},\cdots, \alpha_{i,d} \right]$, and $\bm{x}=[x_1,\cdots, x_d]$.
\end{definition}
\noindent
Here, $L$ is the number of derivatives, $\nabla_{\boldsymbol\alpha_i} y, \; i=1,\cdots,L$ are the $L$ derivatives in the differential operator $\mathcal{F}_{\bm{x}}[y]$, with each a $d$-dimensional multiple index $\boldsymbol\alpha_i\in\mathbb{N}^d$ indicating the order of derivative for each input dimension, and  $c_i(\bm{x})$ is the corresponding coefficient at $\bm{x}\in \mathcal{X}$. 

We refer to the corresponding derivative process $y^\mathcal{F}(\bm{x}): = \mathcal{F}_{\bm{x}}[y](\bm{x})$ as the \textit{PDE process}, and our goal is to find the posterior distribution of $y(\bm{x}_{\rm new})$ given both measurement data $\bm{y}_{1:n}$ and the PDE process $y^\mathcal{F}(\bm{x})$. To this end, we introduce a set of pseudo points $\{\bm{x}^\mathcal{F}_j\}_{j=1}^m$, named \textit{PDE points}, and incorporate the PDE information only at those $m$ points. 
A design method for $\{\bm{x}^\mathcal{F}_j\}_{j=1}^m$ will be discussed later in \cref{sec:APIK}.
Suppose $\{\bm{x}^\mathcal{F}_j\}_{j=1}^m$ are selected, and denote the response vector of the PDE process $\bm{y}^\mathcal{F}_{1:m} =[b(\bm{x}^\mathcal{F}_1),\cdots,b(\bm{x}^\mathcal{F}_m)]^\top$.
We further let $\mathcal{I}[\cdot]$ be the identity operator with $\mathcal{I}_{\bm{x}}[y](\bm{x})=y(\bm{x})$, and then the derivatives of the correlation function and mean function are denoted as
\begin{align}
R_{\boldsymbol\theta}^{\mathcal{F}\mathcal{F}}(\bm{x},\bm{x}') &= (\mathcal{F}_{\bm{x}} \times \mathcal{F}_{\bm{x}'})[R_{\boldsymbol\theta}](\bm{x},\bm{x}') : = \mathcal{F}_{\bm{x}} [ \mathcal{F}_{\bm{x}'}[R_{\boldsymbol\theta}]](\bm{x},\bm{x}'), 
\\
R_{\boldsymbol\theta}^{\mathcal{I}\mathcal{F}}(\bm{x},\bm{x}') &= (\mathcal{I}_{\bm{x}} \times \mathcal{F}_{\bm{x}'})[R_{\boldsymbol\theta}](\bm{x},\bm{x}'):= \mathcal{I}_{\bm{x}} [ \mathcal{F}_{\bm{x}'}[R_{\boldsymbol\theta}]](\bm{x},\bm{x}'),\\
\text{and} \quad \mu^\mathcal{F}(\bm{x}) & =\bm{p}^\top_\mathcal{F}(\bm{x})\boldsymbol\beta.
    \label{eq:DerivativeProcess}
\end{align}
Here, $\bm{p}_\mathcal{F}(\bm{x})=[\mathcal{F}_{\bm{x}}[p_1](\bm{x}),\cdots, \mathcal{F}_{\bm{x}}[p_q](\bm{x})]^\top$ contains the $q$ derivatives of the basis functions, and notation $\mathcal{F}_{\bm{x}} \times \mathcal{F}_{\bm{x}'}$ stands for taking two differential operators $\mathcal{F}_{\bm{x}}[\cdot]$ and $\mathcal{F}_{\bm{x}'}[\cdot]$ on the two inputs of the correlation function $R_{\boldsymbol\theta}(\bm{x},\bm{x}')$, respectively. Note that we select the basis functions $\{p_i(\cdot)\}_{i=1}^q$ and the correlation function $R_{\boldsymbol\theta}(\cdot,\cdot)$ such that the above derivatives exist. Since both covariance operator and differential operator are linear, we have the following proposition for the response vector of measurement points $\bm{y}_{1:n}$ and that of PDE points $\bm{y}^\mathcal{F}_{1:m}$.
\begin{proposition} [Theorem 2.2.2 in \cite{adler1981geometry}]
\label{le:JGP}
Suppose that the Gaussian process model \eqref{eq:DataModel}-\eqref{eq:NGP_model} holds with known model parameters $\{\boldsymbol\beta, \sigma^2, \boldsymbol\theta\}$, and the differential operator $\mathcal{F}_{\bm{x}}[\cdot]$ has a linear form \eqref{eq:LinearPDE}. Then, response vectors $\bm{y}_{1:n}$ and $\bm{y}^\mathcal{F}_{1:m}$ are jointly normal distributed: 
\begin{align}
    \begin{bmatrix}
    \bm{y}_{1:n}  \vspace{0.1cm} \\
    \bm{y}^\mathcal{F}_{1:m}
\end{bmatrix} \sim \mathcal{N}\left(\begin{bmatrix}
    \bm{P}\vspace{0.1cm}\\
    \bm{P}_\mathcal{F}
\end{bmatrix}\boldsymbol\beta , \; \sigma^2 \boldsymbol\Gamma_{\boldsymbol\theta} \right), \quad \text{with} \quad 
\boldsymbol\Gamma_{\boldsymbol\theta} = \begin{bmatrix}
    \bm{R}_{\mathcal{I}\mathcal{I}} & \bm{R}_{\mathcal{I}\mathcal{F}}\vspace{0.1cm}\\
    \bm{R}_{\mathcal{I}\mathcal{F}}^\top  & \bm{R}_{\mathcal{F}\mathcal{F}}
\end{bmatrix}.
\label{eq:jointG}
\end{align}
Here, $\bm{P}_\mathcal{F}= [\bm{p}_\mathcal{F}(\bm{x}^\mathcal{F}_1),\cdots, \bm{p}_\mathcal{F}(\bm{x}^\mathcal{F}_m)]^\top$,
$\bm{R}_{\mathcal{I}\mathcal{I}} = \bm{R}_{\boldsymbol\theta}$, $\bm{R}_{\mathcal{I}\mathcal{F}}={[R_{\boldsymbol\theta}^{\mathcal{I}\mathcal{F}}(\bm{x}_i,\bm{x}^\mathcal{F}_j)]_{i=1}^n}_{j=1}^m$, and  $\bm{R}_{\mathcal{F}\mathcal{F}}={[R_{\boldsymbol\theta}^{\mathcal{F}\mathcal{F}}(\bm{x}_i^\mathcal{F},\bm{x}_j^\mathcal{F})]_{i=1}^n}_{j=1}^m$. 
\end{proposition}
\noindent In words, $\bm{P}_\mathcal{F}$ is the $m$ by $q$ model matrix for PDE points with the derivatives of basis functions from \eqref{eq:DerivativeProcess}, and the $(n+m)$ by $(n+m)$ correlation matrix
$\boldsymbol\Gamma_{\boldsymbol\theta}$ can be written in a 2-by-2 block matrix form, with
$\bm{R}_{\mathcal{I}\mathcal{I}}$ the standard $n$ by $n$ correlation matrix among observations at  $\{\bm{x}_i\}_{i=1}^n$ in \eqref{eq:NGP_E},  $\bm{R}_{\mathcal{F}\mathcal{F}}$ the $m$ by $m$ correlation matrix among the PDE points $\{\bm{x}^\mathcal{F}_j\}_{j=1}^m$ with the correlation function $R_{\boldsymbol\theta}^{\mathcal{F}\mathcal{F}}(\cdot,\cdot)$, and $\bm{R}_{\mathcal{I}\mathcal{F}}$ the $n$ by $m$ correlation matrix between $\{\bm{x}_i\}_{i=1}^n$ and $\{\bm{x}^\mathcal{F}_j\}_{j=1}^m$  with the correlation function $R_{\boldsymbol\theta}^{\mathcal{I}\mathcal{F}}(\cdot,\cdot)$.

Now, let $\tilde{\bm{y}}_{n+m} = [\bm{y}_{1:n}^\top,\bm{y}^\mathcal{F~\top}_{1:m}]^\top$ be an $(n+m)$-vector concatenating the response vectors of measurement points and PDE points. Combining
\cref{le:NGP,le:JGP}, we have the following theorem for the posterior distribution of the PIK model with linear PDEs.
\begin{theorem1}
\label{prop:PIKlinear}
Suppose that the Gaussian process model \eqref{eq:DataModel}-\eqref{eq:NGP_model} holds with known model parameters $\{\boldsymbol\beta, \sigma^2, \boldsymbol\theta\}$, and the differential operator  $\mathcal{F}_{\bm{x}}[\cdot]$ has a linear form \eqref{eq:LinearPDE}.
Then, given the response vector $\tilde{\bm{y}}_{n+m}$, the posterior distribution $y(\bm{x}_{\rm new})$ at a new input location $\bm{x}_{\rm new}$ is a normal distribution with mean and variance 
\begin{align}
\hat{y}(\mathbf{x}_{\rm new}) &:=\mathbb{E} \left[y(\bm{x}_{\rm new})|\tilde{\bm{y}}_{n+m}\right]  =  \bm{p}^\top(\bm{x}_{\rm new})\boldsymbol\beta + \boldsymbol\gamma^\top(\bm{x}_{\rm new}) \boldsymbol\Gamma_{\boldsymbol\theta}^{-1}(\tilde{\bm{y}}_{n+m}-\tilde{\bm{P}}\boldsymbol\beta),
\label{eq:GPPDE_E}\\
s^2(\mathbf{x}_{\rm new}) &:= \text{\rm Var} \left[y(\bm{x}_{\rm new})|\tilde{\bm{y}}_{n+m}\right]  =  \sigma^2 (1- \boldsymbol\gamma^\top(\bm{x}_{\rm new})\boldsymbol\Gamma_{\boldsymbol\theta}^{-1}\boldsymbol\gamma(\bm{x}_{\rm new})), \label{eq:GPPDE_V}
\end{align}
with $\tilde{\bm{P}}= [\bm{P}^\top,\bm{P}_\mathcal{F}^\top]^\top$ and $\boldsymbol\gamma(\bm{x}) = [\bm{r}^\top(\bm{x}),R_{\boldsymbol\theta}^{\mathcal{I}\mathcal{F}}(\bm{x},\bm{x}^\mathcal{F}_1),\cdots,R_{\boldsymbol\theta}^{\mathcal{I}\mathcal{F}}(\bm{x},\bm{x}^\mathcal{F}_m)]^\top$.  
\end{theorem1}
\noindent
Here, $\tilde{\bm{P}}$ is the $(n+m)$ by $q$ model matrix for measurement points and PDE points, $\boldsymbol\Gamma _{\boldsymbol\theta}$ is the correlation matrix defined in \eqref{eq:jointG}, and $\boldsymbol\gamma(\bm{x})$ is the corresponding correlation vector. The proof of \cref{prop:PIKlinear} can be found in \cref{app:PIKlinear}, which can be derived from the conditional normal formulation.

With linear PDEs, \cref{prop:PIKlinear} indicates that the proposed PIK model  can elegantly incorporate the PDE information into the Gaussian process probabilistic modeling framework, via a set of PDE points similar to the measurement data. 
Despite more cumbersome notation,  both conditional mean \eqref{eq:GPPDE_E} and conditional variance \eqref{eq:GPPDE_V} enjoy the closed-form expressions, which can be efficiently computed via the measurement data and the PDEs at hand. Similar to the standard kriging method, the conditional mean can be used for prediction, and the conditional variance can be used for constructing point-wise predictive intervals.

We are also interested in the predictive distribution of the derivative of the physical quantity of interest, {\it i.e.}, $y^{\mathcal{G}}(\bm{x}_{\rm new}) := \mathcal{G}_{\bm{x}}[y](\bm{x}_{\rm new})$, where $\mathcal{G}_{\bm{x}}[\cdot]$ is a linear differential operator \eqref{eq:LinearPDE}. We have the following corollary for the posterior prediction.

\begin{corollary}
\label{cor:linearPDE}
With the same model in \cref{prop:PIKlinear}, the posterior distribution of a derivative process $y^{\mathcal{G}}(\bm{x}_{\rm new})|\tilde{\bm{y}}_{n+m}$ at a new input location $\bm{x}_{\rm new}$ is a normal distribution with
\begin{align}
\mathbb{E} \left[y^{\mathcal{G}}(\bm{x}_{\rm new})|\tilde{\bm{y}}_{n+m}\right] & =  \bm{p}_\mathcal{G}^\top(\bm{x}_{\rm new})\boldsymbol\beta + \boldsymbol\gamma_\mathcal{G}^\top(\bm{x}_{\rm new}) \boldsymbol\Gamma_{\boldsymbol\theta}^{-1}(\tilde{\bm{y}}_{n+m}-\tilde{\bm{P}}\boldsymbol\beta),
\label{eq:GPPDE_der_E} \\
\text{\rm Var}\left[y^{\mathcal{G}}(\bm{x}_{\rm new})|\tilde{\bm{y}}_{n+m}\right] & =  \sigma^2 (r_{\mathcal{G}}(\bm{x})- \boldsymbol\gamma_\mathcal{G}^\top(\bm{x}_{\rm new})\boldsymbol\Gamma_{\boldsymbol\theta}^{-1}\boldsymbol\gamma_\mathcal{G}(\bm{x}_{\rm new})),
\label{eq:GPPDE_der_V}
\end{align}
where $\bm{p}_\mathcal{G}(\bm{x})=[\mathcal{G}_{\bm{x}}[p_1](\bm{x}),\cdots, \mathcal{G}_{\bm{x}}[p_q](\bm{x})]^\top$,   $\boldsymbol\gamma_\mathcal{G}(\bm{x}) = [R_{\boldsymbol\theta}^{\mathcal{G}\mathcal{I}}(\bm{x},\bm{x}_1),\cdots, R_{\boldsymbol\theta}^{\mathcal{G}\mathcal{I}}(\bm{x},\bm{x}_n), \break R_{\boldsymbol\theta}^{\mathcal{G}\mathcal{F}}(\bm{x},\bm{x}^\mathcal{F}_1),\cdots,R_{\boldsymbol\theta}^{\mathcal{G}\mathcal{F}}(\bm{x},\bm{x}^\mathcal{F}_m)]^\top$, and $r_{\mathcal{G}}(\bm{x}) = R_{\boldsymbol\theta}^{\mathcal{G}\mathcal{G}}(\bm{x},\bm{x})$.
\end{corollary}
\noindent
Here, $\bm{p}_\mathcal{G}(\bm{x})$ consists of the derivatives of the $q$ basis functions via $\mathcal{G}_{\bm{x}}[\cdot]$, and  $\boldsymbol\gamma_\mathcal{G}(\bm{x})$ is the correlation vector of the considered derivative process $\mathcal{G}_{\bm{x}}[y]$ and data $\tilde{\bm{y}}_{n+m}$; The notation of the derivatives follow from \eqref{eq:DerivativeProcess}.
The proof of \cref{cor:linearPDE} can be found in \cref{app:linearPDECor}. \cref{cor:linearPDE} shows that, given the response vectors, the posterior distribution of a derivative process at a new point also follows a normal distribution. We will use both conditional mean \eqref{eq:GPPDE_der_E} and variance \eqref{eq:GPPDE_der_V} for in model estimation with nonlinear PDEs (see \cref{sec:learnAPIK}).

\subsection{PIK with nonlinear PDEs} 
\label{sec:nonlinearPDE}

Now we extend the above PIK model for \textit{nonlinear} PDEs.  Here, we consider the following class of nonlinear PDEs.
\begin{definition}[Nonlinear PDE]
$\mathcal{F}_{\bm{x}}[y](\bm{x}) = b(\bm{x})$ is called nonlinear PDE if the differential operator $\mathcal{F}_{\bm{x}}[\cdot]$ is a nonlinear differential operator:  \begin{align}
   \mathcal{F}_{\bm{x}}[y](\bm{x}) =\sum_{i=1}^L c_{i}(\bm{x})\prod_{j=1}^{l_i} \nabla_{\boldsymbol\alpha_{ij}} y(\bm{x}), ~~ \text{where}~~ \boldsymbol\alpha_{ij} = \left[\alpha_{ij,1},\cdots, \alpha_{ij,d} \right]^\top ~\textrm{and}~~ \prod_{i=1}^L l_i > 1.
  \label{eq:NonlinearPDE}
\end{align}
\end{definition}
\noindent
The nonlinear operator in \eqref{eq:NonlinearPDE} consists of $L$ terms with each containing a product of $l_i\geq 1$ derivatives, and at least one of $l_i$'s is strictly greater than one, making the operator $\mathcal{F}_{\bm{x}}[\cdot]$ nonlinear (see an example below). 
Our goal is to find the posterior distribution of $y(\bm{x}_{\rm new})$ given both the measurement data and the derivative process outputs at PDE points $\tilde{\bm{y}}_{n+m}$.
Due to the existence of the \textit{nonlinear} terms in \eqref{eq:NonlinearPDE}, the PDE process $y^\mathcal{F}(\bm{x}) = \mathcal{F}_{\bm{x}}[y](\bm{x})$ is \textit{not} a Gaussian process. This compromises the elegant kriging framework in \cref{prop:PIKlinear}. 
However, by assuming some derivatives are \textit{known}, the ``conditional" differential operator would be still  \textit{linear}.

One example for nonlinear PDEs is the inviscid Burger's equation  \cite{burgers1948mathematical}, which we will look into later in \cref{sec:InBEquation}; It contains the following nonlinear differential operator 
\begin{align}
\mathcal{F}_{\bm{x}}[y]:= \frac{\partial y }{\partial t} + y\frac{\partial y }{\partial z}.
\end{align}
Here, $\bm{x}=[t,z]^\top$ is the input with dimension $d=2$, and the coefficients are $c_1=c_2=1$. The $L=3$ considered derivatives are $\frac{\partial y }{\partial t}$, $y$, and $\frac{\partial y }{\partial z}$ with the corresponding orders $\boldsymbol\alpha_{11} = [1,0]$, $\boldsymbol\alpha_{21} = [0,0]$, and $\boldsymbol\alpha_{22} = [0,1]$, respectively. The nonlinearity is due to the interaction term of $y$ and $\frac{\partial y }{\partial z}$. However, assuming the derivative $\frac{\partial y}{\partial z} = f$ is known, the conditional differential operator $\mathcal{F}_{\bm{x}}[y]|_{\partial y/\partial z=f}= \frac{\partial y}{\partial t} + f y$ becomes a linear differential operator. We can then obtain the posterior mean and variance similar to the PIK method with linear PDEs as in \eqref{eq:GPPDE_E} and \eqref{eq:GPPDE_V}.

With the above intuition in mind, we first select a subset of the derivatives $\{\bar{\boldsymbol\alpha}_k\}_{k=1}^K\subset \{\boldsymbol\alpha_{ij}\}$ with the minimal cardinality $K<\sum_i l_i$, such that one can linearize the differential operator by constructing a new \textit{conditional} operator $\bar{\mathcal{F}}[\cdot]=\mathcal{F}[\cdot]|\partial[\cdot]$ with $\partial =\{ \nabla_{\bar{\boldsymbol \alpha}_1},\cdots,\nabla_{\bar{\boldsymbol \alpha}_K}\}$. Such a linearization is always possible with the considered nonlinear PDEs \eqref{eq:NonlinearPDE}. Furthermore, denote the values of those derivatives $\bar{\boldsymbol\alpha}_k$ ($K$ in total) at the PDE points $\{\bm{x}^\mathcal{F}_j\}_{j=1}^m$ as $ \bm{z} ={\{\partial[y](\bm{x}_j^\mathcal{F})\}_{j=1}^m}$; We then treat $ \bm{z} \in \mathbb{R}^{mK}$ as latent variables. The posterior distribution of $\bm{z}$ would be learned together with the model parameters via an expectation-maximization method (see \cref{sec:learnAPIK}).
We have the following theorem for the posterior prediction of the PIK model with nonlinear PDEs.
\begin{theorem1}
\label{prop:PIKnonlinear}
Suppose that the Gaussian process model \eqref{eq:DataModel}-\eqref{eq:NGP_model} holds with known model parameters $\{\boldsymbol\beta, \sigma^2, \boldsymbol\theta\}$, and the nonlinear differential operator $\mathcal{F}_{\bm{x}}[\cdot]$ has the form  \eqref{eq:NonlinearPDE}. Further assume that $\bar{\bm{z}} := \mathbb{E}[\bm{z}|\tilde{\bm{y}}_{n+m}]$ and $\text{\rm Var}[\bm{z}|\tilde{\bm{y}}_{n+m}]$ are known. 
Then, at a new input location $\bm{x_{\rm new}}$, we have the following posterior mean $\hat{y}(\mathbf{x}_{\rm new})$ and variance $s^2(\mathbf{x}_{\rm new})$:
\begin{align}
\hat{y}(\mathbf{x}_{\rm new}) &=\bm{p}^\top(\bm{x}_{\rm new})\boldsymbol\beta + \boldsymbol\gamma^\top(\bm{x}_{\rm new}) \bar{\boldsymbol\Gamma}_{\boldsymbol\theta}^{-1}([\tilde{\bm{y}}^\top_{n+m},\bar{\bm{z}}^\top ]^\top-\tilde{\bm{P}}\boldsymbol\beta),
\label{eq:APIK_E}
\\
s^2(\mathbf{x}_{\rm new}) &=  \sigma^2 -\sigma^2  \boldsymbol\gamma^\top(\bm{x}_{\rm new})\bar{\boldsymbol\Gamma}_{\boldsymbol\theta}^{-1}\boldsymbol\gamma(\bm{x}_{\rm new})+\boldsymbol\gamma^\top(\bm{x}_{\rm new})\bar{\boldsymbol\Gamma}_{\boldsymbol\theta}^{-1}\boldsymbol\Sigma_{\bm{z}} \bar{\boldsymbol\Gamma}_{\boldsymbol\theta}^{-1}\boldsymbol\gamma(\bm{x}_{\rm new}).
\label{eq:APIK_V}
\end{align}
Here, $\tilde{\bm{P}}= [\bm{P}^\top,\bm{P}_{\bar{\mathcal{F}}}^\top,\bm{P}_{\partial}^\top]^\top$, $\boldsymbol\gamma(\bm{x}) = [\bm{r}^\top(\bm{x}),R_{\boldsymbol\theta}^{\mathcal{I}\bar{\mathcal{F}}}(\bm{x},\bm{x}^\mathcal{F}_{1:m}),R_{\boldsymbol\theta}^{\mathcal{I}{\partial}}(\bm{x},\bm{x}^\mathcal{F}_{1:m})]^\top$, the 3-by-3 block matrix $\bar{\boldsymbol\Gamma}_{\boldsymbol\theta} = [\bm{R}_{\mathcal{I}\mathcal{I}}, \bm{R}_{\mathcal{I}\bar{\mathcal{F}}}, \bm{R}_{\mathcal{I}{\partial}}; \bm{R}_{\mathcal{I}\bar{\mathcal{F}}}^\top,  \bm{R}_{\bar{\mathcal{F}}\bar{\mathcal{F}}},\bm{R}_{\bar{\mathcal{F}}\partial}; \bm{R}^\top_{\mathcal{I}{\partial}},\bm{R}^\top_{\bar{\mathcal{F}}\partial}, \bm{R}_{\partial\partial}]$ with its explicit expression in \cref{app:PIKnonlinear}, and $\boldsymbol\Sigma_{\bm{z}} =  \text{\rm diag}(\bm{0}_{n+m},\text{\rm Var}[\bm{z}|\tilde{\bm{y}}_{n+m}])$.
\end{theorem1}
\noindent In words, $\tilde{\bm{P}}$ is the $(n+m+mK) \times q$ model matrix, concatenating basis functions for the measurement points, the corresponding basis functions under linearized differential operator $\bar{\mathcal{F}}$ at PDE points, and those under latent derivatives $\partial$ also at PDE points, vector $\boldsymbol\gamma(\bm{x})$ is the correlation vector combining the same three parts, and $\bar{\boldsymbol\Gamma}_{\boldsymbol\theta}$ is the  $(n+m+mK) \times (n+m+mK)$ corresponding correlation matrix following from \eqref{eq:jointG}. The proof of \cref{prop:PIKnonlinear} can be found in \cref{app:PIKnonlinear}; Note that the law of total expectation and the law of total variance are used for the posterior mean and variance, respectively. We have overloaded some of the notation from \eqref{eq:GPPDE_E} and \eqref{eq:GPPDE_V}; The difference should be clear from the content. 

According to \cref{prop:PIKnonlinear}, the computation of the posterior mean with \textit{nonlinear} PDEs \eqref{eq:APIK_E} can be viewed as first evaluating the expected value of the latent variable $\bar{\bm{z}}$ and then plugging it into the prediction formulation with \textit{linear} PDEs  \eqref{eq:GPPDE_E}.
For the posterior variance \eqref{eq:APIK_V}, we notice an additional variance term $\boldsymbol\gamma^\top(\bm{x}_{\rm new})\boldsymbol\Gamma_{\boldsymbol\theta}^{-1}\boldsymbol\Sigma_{\bm{z}} \boldsymbol\Gamma_{\boldsymbol\theta}^{-1}\boldsymbol\gamma(\bm{x}_{\rm new})$ with the posterior variance of the latent variables $\boldsymbol\Sigma_{\bm{z}} = \text{diag}(\bm{0}_{n+m},\text{Var}[\bm{z}|\tilde{\bm{y}}_{n+m}])$, compared to that with linear PDEs \eqref{eq:GPPDE_V}. This can be interpreted as accounting for the additional uncertainty due to the variability of the latent $\bm{z}$. Similarly, the conditional mean in \eqref{eq:APIK_E} can be used for prediction, and the conditional variance in \eqref{eq:APIK_V} can be used for quantifying predictive uncertainty.

\subsection{Difference to numerically solving PDEs} 
\label{sec:solvePDE}
There are two major differences between the proposed PIK method and the numerical methods for solving PDEs, {\it e.g.}, finite difference methods \cite{strikwerda2004finite} and finite element methods \cite{johnson2012numerical}. First, 
those numerical methods \textit{require} boundary conditions and initial conditions along with the governing PDEs. In many real applications, however, the boundary conditions are \textit{difficult} to obtain. Take the previous wafer heating application as an example. The boundary condition of temperature is difficult to specify since the wafer is placed on a rotating platform with unknown temperature and complex heat flux. 
Another example is to understand the flood flow in healthcare applications \cite{qian2017quantitative,chen2020active}. While the governing PDEs are known, {\it i.e.}, the Naiver-Stocks equation, the boundary conditions are extremely difficult to obtain, considering the patient-specific blood vessel geometry and the interaction between blood flow and soft biological tissues.
In contrast, the proposed PIK method can utilize measurement points together with PDEs, and provide a prediction of the physical field \textit{without} boundary conditions.

Second, the PIK method provides a natural way to quantify the uncertainty via point-wise predictive intervals, similar to the standard kriging method.
This is because PIK assigns a probabilistic model on the target physical field and then finds the posterior prediction conditional on both measurement data and PDEs. In contrast, standard numerical methods typically adopt a finite discretization. Therefore, the solution is deterministically computed from the boundary to the field without underestimating the associated uncertainty. This is particularly true, when only part of the governing PDE system is known. In such cases, standard numerical methods may \textit{not} provide any results since the physical field is under-determined. A simple example is that one wants to understand a flow field with velocities of two directions while only having one PDE. Nevertheless, our PIK method can still \textit{provide} a prediction with uncertainty quantification of the physical field via posterior distributions. 

\section{APIK model}
\label{sec:APIK}

This section presents the APIK model, which \textit{actively} designs PDE points $\{\bm{x}^\mathcal{F}_j\}_{j=1}^m$ based on the PIK model and the existing measurements $\{\bm{x}_i\}_{i=1}^n$ (see \cref{fig:ill}). We first discuss the design criterion and then present a heuristic method for selecting the PDE data size $m$.

\subsection{Design PDE points}
We propose to select the input locations of the PDE points by minimizing the following integral mean-squared error (IMSE) criterion \cite{sacks1989design,santner2018design}:
\begin{align}
\{\hat{\bm{x}}^\mathcal{F}_j\}_{j=1}^m  &= \argmin_{\bm{x}^\mathcal{F}_1,\cdots,\bm{x}^\mathcal{F}_m}\int_{\mathcal{X}} \text{Var} \left[y(\bm{x}_{\rm new})|\tilde{\bm{y}}_{n+m}\right] \; d\bm{x}_{\rm new}.
\label{eq:IMSE}
\end{align}
IMSE design criterion \eqref{eq:IMSE} can be interpreted as follows. The term $ \text{Var} \left[y(\bm{x}_{\rm new})|\tilde{\bm{y}}_{n+m}\right]$ quantifies the posterior variance ({\it i.e.}, mean-squared error, MSE) of the physical quantity at an untested input $\bm{x}_{\rm new}$, given both measurement data and the \textit{potential} PDE points.
An integral is then taken to find the average predictive uncertainty over the whole input space.
Finally, we select the PDE points $\{\bm{x}^\mathcal{F}_j\}_{j=1}^m$ by minimizing  IMSE \eqref{eq:IMSE} to ensure the designed PDE points can yield low predictive uncertainty and better predictive accuracy.  

The selected PDE points by \eqref{eq:IMSE} can leverage the PDE information in two aspects. First, we explicitly minimize the predictive uncertainty for the potential PIK models. To this end, the selected PDE points not only improve the predictive accuracy of the learned PIK model but also provide a better quantification of uncertainty with narrow predictive intervals. Second, IMSE criterion \eqref{eq:IMSE} considers the predictive uncertainty conditional on \textit{both} measurement data and PDE data, to explore their correlation. 
This ensures that the selected PDE locations minimize the predictive uncertainty according to the measurement data. 
As a result, our APIK model leverages the PDE information via a carefully designed set of PDE points, and therefore, can be shown to achieve better learning performance.

\subsection{PDE data size}\label{sec:APIKmsize}
We present a heuristic method for selecting the size of the PDE points. For simplicity, we  consider only linear PDEs in this subsection. As for the nonlinear PDEs, we suggest using the corresponding linearized PDEs in \cref{sec:nonlinearPDE}. 
We first define the variance reduction ratio (VRR) of the process $y(\cdot)$, with input locations $\{\bm{x}_i\}_{i=1}^n$ and the corresponding responses $\bm{y}_{1:n}$:
\begin{align}
    \text{\rm VRR}(\{\bm{x}_i\}_{i=1}^n;y(\cdot)) := \frac{\int_{\mathcal{X}} \text{\rm Var} \left[y(\bm{x}_{\rm new})\right] d\bm{x}_{\rm new} - \int_{\mathcal{X}} \text{\rm Var} \left[y(\bm{x}_{\rm new})|\bm{y}_{1:n}\right] d\bm{x}_{\rm new}}{\int_{\mathcal{X}} \text{\rm Var} \left[y(\bm{x}_{\rm new})\right] d\bm{x}_{\rm new}} .
    \label{eq:VR}
\end{align}
Given a correlation function $R_{\boldsymbol\theta}(\cdot,\cdot)$ for $y(\cdot)$, we have the following proposition for VRR.

\begin{proposition}[Theorem 1 in \cite{harari2018computer}]
\label{le:IMSEMeansure} Given budget $n$ (i.e., $n$ measurement points), 
\begin{align}
\sup_{\bm{x}_i, \cdots, \bm{x}_n \in \mathcal{X}} \text{\rm VRR}(\{\bm{x}_i\}_{i=1}^n;y(\cdot)) \leq\frac{\sum_{i=1}^n \lambda_{i}}{\sum_{i=1}^\infty \lambda_{i}}, 
\end{align}
where $\lambda_1\geq\lambda_2\geq\cdots\geq0$ are the eigenvalues of the correlation function $R_{\boldsymbol\theta}(\bm{x},\bm{x}')$. 
\end{proposition}
\noindent
\cref{le:IMSEMeansure} indicates that in the optimal case, the variance reduction due to the $n$ measurement points $\{\bm{x}_i\}_{i=1}^n$ would be the summation of the top $n$ eigenvalues $\sum_{i=1}^n\lambda_i$. For the PDE points, we have the following theorem describing a similar bound for the \textit{derivative} process $y^\mathcal{F}(\cdot)$ with PDE points.
\begin{theorem1}
\label{prop:IMSEPDE} 
Suppose that the Gaussian process model \eqref{eq:DataModel}-\eqref{eq:NGP_model} holds and the differential operator  $\mathcal{F}_{\bm{x}}[\cdot]$ has a linear form \eqref{eq:LinearPDE}. Moreover, the correlation function is sufficiently smooth, i.e., $(\mathcal{F}_{\bm{x}}\times \mathcal{F}_{\bm{x}'})[R_{\boldsymbol\theta}](\bm{x},\bm{x}')$ exists. Then the correlation function of $y^\mathcal{F}(\cdot)$ has the expansion:  
\begin{align}\label{eq:de_comp}
R_{\boldsymbol\theta}^{\mathcal{F}\mathcal{F}}(\bm{x},\bm{x}') =\sum_{i=1}^\infty \xi_i \varepsilon_i(\bm{x})\varepsilon_i(\bm{x}'),    
\end{align}
where $\xi_1\geq\xi_2\geq \cdots \geq 0$ are eigenvalues and $\varepsilon_i(\cdot)$'s are the corresponding eigenfunctions. Furthermore, given $m$ PDE points and no measurement points, 
\begin{align}
\sup_{\bm{x}^\mathcal{F}_1,\cdots \bm{x}^\mathcal{F}_m\in\mathcal{X}^m} \text{\rm VRR}(\{\bm{x}^\mathcal{F}_j\}_{j=1}^m; y^\mathcal{F}(\cdot)) \leq \frac{\sum_{j=1}^m\xi_j}{\sum_{j=1}^\infty\xi_j}.
\end{align}
\end{theorem1}
\noindent
Here, instead of the correlation function for the physical quantity $R_{\boldsymbol\theta}(\cdot,\cdot)$, we decompose the correlation function for the PDE process $R_{\boldsymbol\theta}^{\mathcal{F}\mathcal{F}}(\cdot,\cdot)$ by Mercer's theorem. Similarly, $\{\xi_j\}_{j=1}^\infty$ is the set of eigenvalues with $\xi_1 \geq \xi_2\geq\cdots\geq 0$ and $\{\varepsilon_j(\bm{x})\}_{j=1}^\infty$ is the set of corresponding orthonormal eigenfunctions in $L^2(\mathcal{X})$. 
\cref{prop:IMSEPDE} essentially says that the variance reduction by PDE points can also be bounded by the eigenvalues of the correlation function for the PDE process;
A formal proof of this theorem can be found in \cref{app:IMSEPDE}.

According to \cref{le:IMSEMeansure} and \cref{prop:IMSEPDE}, the variance reduction due to either measurement points or PDE points can be quantified by the corresponding eigenvalues.
To ensure that the set of PDE points $\{\bm{x}^\mathcal{F}_j\}_{j=1}^m$ \textit{does} contribute to the modeling while \textit{not} overwhelm the information of the measurement data $\{\bm{x}_i\}_{i=1}^n$, we propose to select $m$ such that the VRRs of the two parts are similar:
\begin{align}
m^*=\argmin_{m\in \mathbb{N}}\left[
  \frac{\sum_{i=1}^n \lambda_i}{\sum_{i=1}^\infty \lambda_i} -  \frac{\sum_{j=1}^m \xi_j}{\sum_{j=1}^\infty \xi_j}\right]^2.
  \label{eq:selectm}
\end{align}
The proposed selection method is similar to selecting leading components in principal components analysis: We first compute the eigenvalues of the two correlation functions $R_{\boldsymbol\theta}(\cdot,\cdot)$ and $R_{\boldsymbol\theta}^{\mathcal{F}\mathcal{F}}(\cdot,\cdot)$, and then compare the two eigenvalue ratios and find the optimal $m$.

\section{APIK model estimation}
\label{sec:learnAPIK}
We present 
an expectation-maximization (EM) method for estimating the model parameters $\boldsymbol\phi = \{\boldsymbol\theta, \boldsymbol \beta,\sigma^2\}$ and the posterior moments of the latent variables $\bm{z}$, and actively selecting the PDE points $\{\bm{x}_j^\mathcal{F}\}_{j=1}^m$. Specifically, we fit a kriging model with only measurement data, and initialize the posterior distribution of the latent $\bm{z}$ by its posterior distribution in \eqref{eq:GPPDE_der_E} and \eqref{eq:GPPDE_der_V}. We initialize the PDE points $\{\bm{x}_j^\mathcal{F}\}_{j=1}^m$ by an $m$-run Sobol' sequence \cite{sobol1967distribution}. We then iterate the following three steps: (i) Expectation step -- given the PDE points $\{\bm{x}_j^\mathcal{F}\}_{j=1}^m$, we take the expectation of the full likelihood with respect to the posterior distribution of $\bm{z}$; (ii) Maximization step -- we optimize the expected likelihood and find the model parameters $\boldsymbol\phi$; (iii) Design step -- given $\boldsymbol\phi$ and $\bm{z}$, we update $\{\bm{x}_j^\mathcal{F}\}_{j=1}^m$ via minimizing the IMSE criterion. \cref{alg:nonlinearAPIK} summarizes the iterative method, with each step to be discussed in detail in this section.
Note that the following discussion is for the APIK model with nonlinear PDEs \eqref{eq:NonlinearPDE}; If the considered PDEs are linear \eqref{eq:LinearPDE}, we discard the expectation step and iterate over the maximization step and the design step.

\begin{algorithm}[t]
\caption{\texttt{LearnAPIK}($\bm{y}_{1:n},\mathcal{F}_{\bm{x}}[\cdot],\text{MaxIter}$): Estimate APIK with nonlinear PDEs}\label{alg:nonlinearAPIK}
\begin{algorithmic}[1]
\small
\stb obtain the size of PDE points $m$ via \eqref{eq:selectm}
\stb initialize $\{\bm{x}_j^\mathcal{F}\}_{j=1}^m \leftarrow  \texttt{Sobol sequence}(m,p)$
\stb set $b_j \leftarrow b(\bm{x}_j^\mathcal{F})$ for $j=1,2,\cdots, m$ and set $\tilde{\bm{y}}_{n+m}\leftarrow [\bm{y}_{1:n}^\top,b_1, \cdots, b_m]^\top$
\stb generate Monte Carlo input samples $\{\bm{x}^{MC}_k\}_{k=1}^{100n}$ in the input domain $\mathcal{X}$
\For {$k=1,\cdots,\text{MaxIter}$}
\For {$j=1,\cdots,m$}
\stb set $\bar{\bm{z}}\leftarrow \mathbb{E}(
\bm{z}|\tilde{\bm{y}}_{n+m})$  by \eqref{eq:APIK_E}
\stb set $\boldsymbol\Sigma_{\bm{z}} \leftarrow \text{diag}(\bm{0}_{n+m},\text{Var}[\bm{z}|\tilde{\bm{y}}_{n+m};\boldsymbol\phi])$  by \eqref{eq:APIK_V}
\stb set function $Q(\boldsymbol\theta,\boldsymbol\beta,\sigma^2) \leftarrow \mathbb{E}\left[l(\boldsymbol\phi ;\tilde{\bm{y}}_{n+m},\bm{z})|\tilde{\bm{y}}_{n+m},\boldsymbol\phi\right]$ by \eqref{eq:pluginE} \Comment{Expectation step}
\Function{ENLL}{$\boldsymbol\theta$}
\stb obtain $\boldsymbol\beta$ and  $\sigma^2$ by \eqref{eq:NonlinearFOC} using $\boldsymbol\theta$
\stb \Return $Q(\boldsymbol\theta,\boldsymbol\beta,\sigma^2)$
\EndFunction
\stb optimize $\boldsymbol\theta^* \leftarrow \min \text{ENLL}(\boldsymbol\theta)$ by L-BFGS \Comment{Maximization step}
\stb update $\boldsymbol\beta$ and  $\sigma^2$ by \eqref{eq:NonlinearFOC} using $\boldsymbol\theta^*$ 
\stb set $\boldsymbol\Gamma_{\boldsymbol\theta}$ by    \eqref{eq:APIK_E}, and set $\boldsymbol\Gamma_{\boldsymbol\theta-j} \leftarrow \boldsymbol\Gamma_{\boldsymbol\theta}[-j,-j]$ by deleting the $j$-th row and column 
\stb compute $\boldsymbol\Gamma_{\boldsymbol\theta-j}^{-1}$ by inverting the matrix $\boldsymbol\Gamma_{\boldsymbol\theta-j}$
\Function{IMSE}{$\bm{x}$}
\stb obtain $\boldsymbol\Gamma_{\boldsymbol\theta}^{-1}$ by \eqref{eq:fastInverse}
\stb compute the predictive variance $s^2_k \leftarrow s^2(\bm{x}^{MC}_k)$ via \eqref{eq:APIK_V} for $k=1,\cdots,100n$
\stb \Return the average $ \sum_k(s^2_k)/(100n)$
\EndFunction
\stb optimize $\bm{x}_j^\mathcal{F} \leftarrow \min \text{IMSE}(\bm{x})$ by L-BFGS
\Comment{Design step}
\stb update $b_j \leftarrow b(\bm{x}_j^\mathcal{F})$
\EndFor
\EndFor
\stb \Return $\boldsymbol\phi = \{\boldsymbol\beta,\boldsymbol\theta,\sigma^2\}$, $\{\bar{\bm{z}},\boldsymbol\Sigma_{\bm{z}}\}$, and $\{\bm{x}_j^\mathcal{F}\}_{j=1}^m$
\normalsize
\end{algorithmic}
\end{algorithm}

\subsection{Expectation step}
\label{sec:Estep}
Given the PDE points $\{\bm{x}_j^\mathcal{F}\}_{j=1}^m$, we want to first find the  likelihood for measurement data, PDE process outputs at PDE points, and latent variables $\bm{z}$. While the joint distribution of $[\tilde{\bm{y}}^\top_{n+m},
\bm{z}^\top]^\top$ is \textit{not} Gaussian, we have the full negative log-likelihood $l(\boldsymbol\phi;\tilde{\bm{y}}_{n+m},\bm{z})$ via the \textit{linearized} PDE operator (see detailed derivation in \cref{app:like}):
\begin{align}
      \log( \det(\sigma^2\boldsymbol\Gamma_{\boldsymbol\theta})) +\frac{1}{\sigma^2}\left(\left[\tilde{\bm{y}}^\top_{n+m},
\bm{z}^\top\right]^\top-\tilde{\bm{P}}\boldsymbol\beta\right)^\top\boldsymbol\Gamma_{\boldsymbol\theta}^{-1}\left(\left[\tilde{\bm{y}}^\top_{n+m},
\bm{z}^\top\right]^\top-\tilde{\bm{P}}\boldsymbol\beta\right).
\label{eq:nonlinearLike}
\end{align}
Here, $\boldsymbol\Gamma_{\boldsymbol\theta}$ is the $(n+m+mK)$ by $(n+m+mK)$ correlation matrix, and $\tilde{\bm{P}}$ is the $(n+m+mK)$ by $q$ correlation matrix, following from the PIK model with nonlinear PDEs (see \cref{prop:PIKnonlinear}).
The goal is to compute the expectation of the negative log-likelihood $\mathbb{E}\left[l(\boldsymbol\phi ;\tilde{\bm{y}}_{n+m},\bm{z})|\tilde{\bm{y}}_{n+m},\boldsymbol\phi\right]$ with respect to the posterior distribution for $\bm{z}|\tilde{\bm{y}}_{n+m},\boldsymbol\phi$.

Typically, this expectation should be evaluated via Monte Carlo samples due to the complex likelihood structure -- note that the correlation matrix $\boldsymbol\Gamma_{\boldsymbol\theta}$ is also a function of the latent variables $\bm{z}$ through the linearized differential operator $\bar{\mathcal{F}}[\cdot]$. However, the high dimensionality ({\it i.e.}, $mK$, the number of PDE points by the number of derivatives) of $\bm{z}$ would require thousands of Monte Carlo samples for an effective approximation. This can be quite time-consuming in practice.
To address that, we adopt a partially plugged-in method for evaluating the expectation, which would noticeably speed up the computation.

Specifically, we propose to approximate the expectation by plugging the posterior mean $\bar{\bm{z}} = \mathbb{E} [\bm{z}|\tilde{\bm{y}}_{n+m};\boldsymbol\phi]$ in \textit{only} the correlation matrix $\boldsymbol\Gamma_{\boldsymbol\theta}$ due to the lack of closed-form expression, while computing the \textit{full} expectation for the remaining part, {\it i.e.}, the quadratic term. Therefore, the expectation $\mathbb{E}\left[l(\boldsymbol\phi ;\tilde{\bm{y}}_{n+m},\bm{z})|\tilde{\bm{y}}_{n+m},\boldsymbol\phi\right]$ can be approximated by 
\begin{align}
 \log(\det(\sigma^2\bar{\boldsymbol\Gamma}_\theta))+\frac{1}{\sigma^2}\Big[
 \big([\tilde{\bm{y}}^\top_{n+m},
\bar{\bm{z}}^\top]^\top-\tilde{\bm{P}}\boldsymbol\beta\big)^\top\bar{\boldsymbol\Gamma}_\theta^{-1}\big([\tilde{\bm{y}}^\top_{n+m},
\bar{\bm{z}}^\top]^\top-\tilde{\bm{P}}\boldsymbol\beta\big)+ \text{tr}\big(\bar{\boldsymbol\Gamma}_\theta^{-1}\boldsymbol\Sigma_{\bm{z}}\big)\Big].
\label{eq:pluginE}
\end{align}
Here, 
$\bar{\boldsymbol\Gamma}_\theta$ is the correlation matrix with the expected value $\bar{\bm{z}}$ plugged in, and $\boldsymbol\Sigma_{\bm{z}} =  \text{diag}(\bm{0}_{n+m}, \break \text{Var}[\bm{z}|\tilde{\bm{y}}_{n+m};\boldsymbol\phi])$ is the posterior covariance matrix already defined in \cref{prop:PIKnonlinear}. Compared to the negative log-likelihood in \eqref{eq:nonlinearLike}, we notice an additional trace term $\text{tr}(\bar{\boldsymbol\Gamma}_\theta^{-1}\boldsymbol\Sigma_{\bm{z}})$ in \eqref{eq:pluginE} to account for the variability of the hidden $\bm{z}$. This is summarized in lines 7 - 9 in \cref{alg:nonlinearAPIK}.

\subsection{Maximization step}
\label{sec:Mstep}
The goal of the maximization step is to obtain the model parameters $\boldsymbol\phi = \{\boldsymbol\theta, \boldsymbol \beta,\sigma^2\}$ by minimizing the expectation $\mathbb{E}\left[l(\boldsymbol\phi ;\tilde{\bm{y}}_{n+m},\bm{z})|\tilde{\bm{y}}_{n+m},\boldsymbol\phi\right]$ in \eqref{eq:pluginE}. Note that this step can be further reduced to the optimization of only the correlation parameters $\boldsymbol\theta$, by plugging  the following two first-order conditions in the objective function \eqref{eq:pluginE}
\begin{align}
\begin{split}
    \hat{\boldsymbol\beta} &= \left(\tilde{\bm{P}}^\top 
    \bar{\boldsymbol\Gamma}_\theta^{-1} \tilde{\bm{P}}\right)^{-1}\tilde{\bm{P}}^\top 
    \bar{\boldsymbol\Gamma}_\theta^{-1}  \left[\tilde{\bm{y}}^\top_{n+m},
\bar{\bm{z}}^\top\right]^\top, \quad\\
   \widehat{\sigma^2} &=\frac{1}{N} \left[
 \left(\left[\tilde{\bm{y}}^\top_{n+m},
\bar{\bm{z}}^\top\right]^\top-\tilde{\bm{P}}\hat{\boldsymbol\beta}\right)^\top\bar{\boldsymbol\Gamma}_\theta^{-1}\left(\left[\tilde{\bm{y}}^\top_{n+m},
\bar{\bm{z}}^\top\right]^\top-\tilde{\bm{P}}\hat{\boldsymbol\beta}\right)+ \text{tr}\left(\bar{\boldsymbol\Gamma}_\theta^{-1}\boldsymbol\Sigma_{\bm{z}}\right)\right],
\end{split}
 \label{eq:NonlinearFOC}
\end{align}
where $N=n+m+mK$. Note that the additional trace term in \eqref{eq:pluginE} also appears in the closed-form expression for $ \widehat{\sigma^2}$.
Then we treat these two variables operators of $\boldsymbol \theta$ and minimize $l(\boldsymbol\theta, \hat{\boldsymbol \beta}(\boldsymbol\theta), \widehat{\sigma^2}(\boldsymbol\theta,\hat{\boldsymbol \beta}(\boldsymbol\theta)); \tilde{\bm{y}}_{n+m})$ over $\boldsymbol \theta$ by line search methods, {\it e.g.}, the L-BFGS method \cite{liu1989limited}. This part is summarized in lines 10 - 14 in \cref{alg:nonlinearAPIK}.

\subsection{Design step}
\label{sec:Dstep} 
Given the model parameters $\boldsymbol\phi = \{\boldsymbol\theta, \boldsymbol \beta,\sigma^2\}$, we want to update the PDE points via minimizing IMSE criterion \eqref{eq:IMSE}.
However, optimization \eqref{eq:IMSE} is extremely high-dimensional, containing $m$ PDE points each in $\mathbb{R}^d$. We therefore propose to adopt the sequential design strategy \cite{lam2008sequential,gramacy2015local}. Specifically, 
we update \textit{one} PDE point  $\bm{x}^\mathcal{F}_{j}$ at each iteration by sequentially minimizing IMSE criterion over one point with other PDE points fixed
\begin{align}
\begin{split}
\hat{\bm{x}}^\mathcal{F}_{j}= \argmin_{\bm{x}^\mathcal{F}_{j}} \int_{\mathcal{X}} \text{Var} \left[y(\bm{x}_{\rm new})|\bm{y}_{1:n},\bm{y}^\mathcal{F}_{1:j-1},b(\bm{x}^{\mathcal{F}}_j),\bm{y}^\mathcal{F}_{j+1:m}\right] \; d\bm{x}_{\rm new}.
\end{split}
\label{eq:IterIMSE}
\end{align}
Here, the posterior variance is obtained in \eqref{eq:APIK_V}, and $b(\cdot)$ is the non-homogeneous term of the PDEs, {\it i.e.}, the output of the PDE process. Note that the computation of objective function \eqref{eq:IterIMSE} involves an evaluation of the inverse correlation matrix $\boldsymbol\Gamma_{\boldsymbol\theta}^{-1}$ in the posterior variance $\text{Var}[y(\bm{x}_{\rm new})|\cdot]$.
This can be time-consuming 
since we plan to use line search methods, which require \textit{many} matrix inverse steps for updating \textit{one} PDE point $\bm{x}^{\mathcal{F}}_j$. To address that, we have the following theorem for the efficient computation of $\boldsymbol\Gamma_{\boldsymbol\theta}^{-1}$ in \eqref{eq:IterIMSE}. 
\begin{theorem1}
\label{th:fastupdate}
The inverse $\boldsymbol\Gamma_{\boldsymbol\theta}^{-1}$ can be efficiently computed as follows:
\begin{align}
\boldsymbol\Gamma_{\boldsymbol\theta}^{-1} = 
\left[
\begin{array}{cc}
\boldsymbol\Gamma_{\boldsymbol\theta-j}^{-1} + \bm{g}_j(\bm{x}^{\mathcal{F}}_j)  \bm{g}_j(\bm{x}^{\mathcal{F}}_j)^\top s^2_j(\bm{x}^{\mathcal{F}}_j)   & \bm{g}_j(\bm{x}^{\mathcal{F}}_j) \vspace{0.1cm} \\
 \bm{g}_j(\bm{x}^{\mathcal{F}}_j) ^\top    & 1/s^{2}_j(\bm{x}^{\mathcal{F}}_j) 
\end{array}
\right],
\label{eq:fastInverse}
\end{align}
where $\boldsymbol\Gamma_{\boldsymbol\theta-j}$ is $\boldsymbol\Gamma_{\boldsymbol\theta}$ without $j$-th row and column, $s^2_j(\bm{x}) = \text{\rm Var}[y(\bm{x})|\bm{y}_{1:n},\bm{y}^{\mathcal{F}}_{1:j-1}, \bm{y}^{\mathcal{F}}_{j+1:m}]$, and $\bm{g}_j(\bm{x}) = \boldsymbol\Gamma_{\boldsymbol\theta-j}^{-1}\boldsymbol\gamma_{-j}(\bm{x})/s^2_j(\bm{x})$.
\end{theorem1} 
\noindent
Here, $s^2_j(\bm{x})$ is the predictive variance given $(n+m-1)$ points by \eqref{eq:GPPDE_V}, and  $\boldsymbol\gamma_{-j}(\bm{x})$ is the corresponding correlation vector with $n$ observed points and $(m-1)$ PDE points except for the $j$-th PDE point from \eqref{eq:APIK_E}. The proof of this theorem can be found in \cref{app:fastupdate}.

\cref{th:fastupdate}  provides an efficient way to solve design problem~\eqref{eq:IterIMSE}. Specifically,  $\boldsymbol\Gamma_{\boldsymbol\theta-j}^{-1} $ does \textit{not} depend on the $j$-th point $\bm{x}^{\mathcal{F}}_j$, and therefore, we can compute it ahead of the optimization iteration and use it in the whole optimization iteration. Combining both sequential strategy \eqref{eq:IterIMSE} and fast computation for the inverse \eqref{eq:fastInverse}, we can significantly speed up the design step.  
This part is summarized in lines 15 - 21 in \cref{alg:nonlinearAPIK}.

\section{Applications}
\label{sec:example} We present in this section four applications. We first look into two one-dimensional (1D) synthetic examples with a linear PDE and a nonlinear PDE, respectively. We then apply the proposed APIK method to two real-world case studies: Shock wave development in flow dynamics and laser heating process in wafer manufacturing.

In all four applications, we use the posterior mean $\hat{y}(\bm{x}_{\rm new})$ for prediction, {\it i.e.}, \eqref{eq:GPPDE_E} with linear PDEs and \eqref{eq:APIK_E} with nonlinear PDEs,  and construct point-wise $2\sigma$ predictive intervals (PIs) via  the posterior mean $\hat{y}(\bm{x}_{\rm new})$ and the  posterior variance $s^2(\bm{x}_{\rm new})$, {\it i.e.}, \eqref{eq:GPPDE_V} with linear PDEs and \eqref{eq:APIK_V} with nonlinear PDEs.
To evaluate predictive accuracy of $\hat{y}(\bm{x}_{\rm new})$, we consider the root mean-squared error (RMSE) metric:
\begin{align}
\text{RMSE} = \sqrt{\sum_{i=1}^{n_{\rm test}}\frac{|y(\bm{x}_i)-\hat{y}(\bm{x}_i)|^2}{n_{\rm test}}}.
\end{align}
Here, $\{\bm{x}_i\}_{i=1}^{n_{\rm test}}$ are $n_{\rm test}$ test points.

One way to quantify the performance of the constructed PIs is to use \textit{interval score}  (IS, \cite{gneiting2007strictly}). For a $2\sigma$ PI $[\hat{y}_l,\hat{y}_u]$, IS is defined as the width of the PI added by penalties of violation
\begin{align}  
\text{IS}=  (\hat{y}_u-\hat{y}_l)+\frac{2}{\alpha}(\hat{y}_l-y^*)_++\frac{2}{\alpha}(y^*-\hat{y}_u)_+ ,
\end{align}
where $y^*$ is the ground truth, $(a)_+ = \textrm{max}(a,0)$, and $\alpha=5\%$ is the corresponding type I error of $2\sigma$ PIs, assuming a Gaussian predictive distribution. We then compute the mean interval score (MIS) over the test set $\{\bm{x}_i\}_{i=1}^{n_{\rm test}}$ as a performance metric over the whole input space.

\subsection{1D linear PDE example} \label{sec:1Dlin}
We first illustrate the proposed APIK method with a 1D \textit{linear} PDE. Suppose the physical quantity of interest has the following mean function
\begin{align}
   y(x) = x\sin(11x+2),\quad x\in [0,1], 
\end{align}
with measurement noise variance $\sigma_\epsilon^2=0.05^2$. 
The measurement points are selected via minimax designs \cite{johnson1990minimax}. 
The test set $\{\bm{x}_i\}_{i=1}^{n_{\rm test}}$ contains $n_{\rm test}=500$ equally-spaced points over the input space $\mathcal{X} = [0,1]$. We would compute both RMSE and MIS on $\{\bm{x}_i\}_{i=1}^{n_{\rm test}}$.

Further suppose the known PDE $\mathcal{F}_{{x}}[y](x) =b(x)$ is delivered in the form
\begin{align}
    \mathcal{F}_{{x}}[y](x) = 100 y(x)+\frac{\partial^2 y}{\partial x^2}(x),\quad b(x)=22\cos(11x+2) , \quad \text{and} \quad x\in [0,1].
    \label{eq:1DlinPDE}
\end{align}
One can easily check that the ground truth function $y(x)$ is one solution of the known PDE $\mathcal{F}_{{x}}[y](x) =b(x)$,  by plugging $y(x)$ into \eqref{eq:1DlinPDE}. The PDE data size $m$ is selected based on the size of measurement data $n$, using the method discussed in \cref{sec:APIKmsize}.

\begin{figure}
\centering
\includegraphics[width=0.99\textwidth]{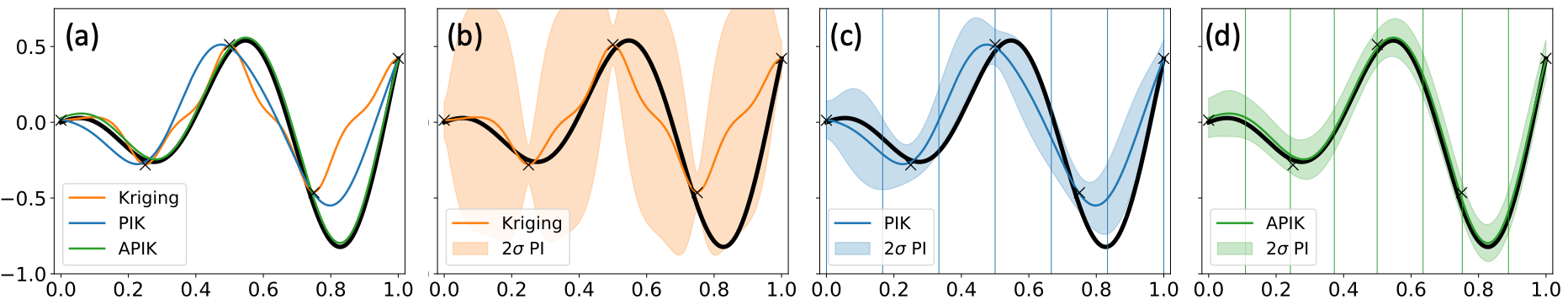}
\caption{\label{fig:L1D} A 1D example with the linear PDE \eqref{eq:1DlinPDE}: A comparison of the standard kriging method (b), the PIK method (c), and the APIK method (d). The solid lines mark the predictive means, the vertical lines mark the locations of PDE points, and the shades outline the quantification of predictive uncertainty.}
\end{figure}

We consider three kriging-based methods, with Gaussian correlation functions $R_\theta({x},{x}') = \exp(-({x}-{x}')^2/\theta)$ and constant mean $\mu({x}) = \beta_0$. The first is the standard kriging method by \textsf{R} package \texttt{DiceKriging} \cite{roustant2010dicekriging}. The second is the proposed PIK model in \cref{sec:PIK} with equally-spaced PDE points. Note that, with \textit{linear} PDEs, the PIK method can be viewed as an extension to the model in \cite{wang2016estimating}. Finally, we consider the proposed APIK model in \cref{sec:APIK}, where the PDE points are actively designed.

\cref{fig:L1D} illustrates the predictions (solid lines) and uncertainty quantification (shades) with $n=5$ measurement points and $m=7$ PDE points, under the three considered methods. As shown in \cref{fig:L1D} (b), the standard kriging method estimates an overly-small scale parameter $\theta$ in the Gaussian correlation function $R_\theta(x,x')$, leading to a clear tread of ``regressing" to the constant mean and large PIs. For PIK with equally-spaced PDE points in \cref{fig:L1D} (c), the prediction is smoother since we also incorporate the PDE information \eqref{eq:1DlinPDE}. However, the predictive accuracy is still not good in some regions ({\it e.g.}, $[0.4,0.6]$) due to the limited data and measurement noise. In addition, while the PIs by PIK are narrower compared to that by standard kriging, they do not cover the true function in input regions of $[0.75,0.9]$.

The proposed APIK method, as shown in \cref{fig:L1D} (d),  performs very well in both prediction and PIs. The posterior mean almost overlaps with the ground truth function, and the PIs cover the true function and are much narrower. This is because APIK leverages the PDE information by selecting PDE points that minimize predictive uncertainty. Particularly, the designed PDE points (vertical lines in \cref{fig:NL1D} (d)) not only explore the whole input domain but also exploit the locations near the measurement points; Those locations appear to provide the most information in reducing the uncertainty. As a result, our APIK model achieves (RMSE, MIS) of (0.08, 0.26), which is much smaller than (0.29, 1.18) for standard kriging and (0.19, 1.05) for PIK.

\begin{figure}[t]
    \vspace{-0.25in}
\noindent\begin{minipage}{0.485\textwidth}
\begin{table}[H]
\resizebox{\columnwidth}{!}{%
\begin{tabular}{c | c | c c c| c c c} 
\toprule
 &  & \multicolumn{3}{c|}{RMSE}   &  \multicolumn{3}{c}{MIS}\\
 \midrule
$n$ & $m$ & Kriging & PIK & APIK & Kriging &PIK & APIK\\
\midrule
4 & 6 & 0.5378   & 0.4509 &  \textbf{0.3244} & 12.232   & 6.092 &   \textbf{3.6122} \\ 
5 & 7 & 0.2904    & 0.1908 & \textbf{0.0842} & 1.179  &  1.048 & \textbf{0.2621} \\
7 & 10 & 0.1708  & 0.1650 &  \textbf{0.0448} & 0.6714   &  0.2415 &\textbf{0.2396}\\
10 & 14 & 0.0504    & 0.0575 & \textbf{0.0386} & 0.2821  &  0.3012 & \textbf{0.2204} \\
15 & 20 & 0.0492 & 0.0409  & \textbf{0.0376} & 0.2213 & 0.2184 &  \textbf{0.2010} \\
\toprule
\end{tabular}
}
\caption{A comparison of the predictive performance (i.e., RMSE and MIS) under the three considered methods in the 1D example with linear PDE \eqref{eq:1DlinPDE}.}
\label{tab:RMSE1DIll}
\end{table}
    \end{minipage}
    \quad
    \begin{minipage}{0.485\textwidth}
\begin{table}[H]
\resizebox{\columnwidth}{!}{%
\begin{tabular}{c | c | c c c| c c c} 
\toprule
 &  & \multicolumn{3}{c|}{RMSE}   &  \multicolumn{3}{c}{MIS}\\
 \midrule
$n$ & $m$ & Kriging & PIK & APIK & Kriging & PIK & APIK\\
\midrule
3 & 5 & 0.4369 &  0.1910 &  \textbf{0.0671} & 0.9058  &  2.7520 & \textbf{0.3521}\\
4 & 6 & 0.0904  &  0.1019 &  \textbf{0.0409} & 0.5111  &  0.5995 & \textbf{0.1985} \\ 
5 & 7 & 0.0565  &  0.0787 & \textbf{0.0292} & 0.2264 &  0.4806 &  \textbf{0.1977}  \\
7 & 10 & 0.0477  & 0.0463 &  \textbf{0.0288} & 0.3974 &  0.2809 & \textbf{0.1825}\\
10 & 14 & 0.0583 &  0.0541 & \textbf{0.0253} & 0.2155  & 0.7376 & \textbf{0.1960} \\
\toprule
\end{tabular}
}
\caption{A comparison of the predictive performance (i.e., RMSE and MIS) under the three considered methods in the 1D example with nonlinear PDE \eqref{eq:1Dnonlin}.}
\label{tab:RMSE1DNonLi}
\end{table}
 \end{minipage}
  \end{figure}

We then consider different measurement data sizes $n\in\{4,5,7,10,15\}$. The corresponding sizes for PDE data are $m\in\{6,7,10,14,20\}$ by \eqref{eq:selectm}.
\cref{tab:RMSE1DIll} (left) shows the RMSEs using the three methods. First, we notice clear improvements of PIK in accuracy compared to standard kriging. This is again because of the incorporation of PDE information. Second, adopting APIK further improves the performance, since APIK not only incorporates the PDE information but also actively designs the PDE points to leverage the PDE information according to measurement points. 
As a result, we notice at least a $20\%$ reduction in RMSE compared to standard kriging. 
Finally, the improvement of APIK is particularly noticeable when the measurement data size is small, which aligns with target small data challenge in engineering applications.
We see over $70\%$ improvements when data size $n\leq 7$.

\cref{tab:RMSE1DIll} (right) shows the MISs using the three considered methods with different data sizes. 
We notice that while PIK struggles in reducing MISs under several measurement sizes ({\it e.g.}, $n=5$ and $10$),  APIK consistently improves the uncertainty quantification performance with smaller MISs. 
This is because APIK actively introduces PDE information by selecting the PDE points with minimal predictive uncertainty. With improvements in both predictive accuracy and uncertainty quantification, the APIK model better incorporates the known PDE information into the kriging framework, and therefore, improves the learning performance with limited measurement data.

\subsection{1D nonlinear PDE example} \label{sec:1Dnonlinex}
Now, consider a synthetic 1D example with a \textit{nonlinear} PDE. Suppose the mean physical quantity of interest and the associated PDE are
\begin{align}
    y(x)=\sin(5x+0.35),  \quad \text{and} \quad y\frac{\partial y}{\partial x}= 2.5\sin(10x+0.7) , \quad x\in [0,1].
    \label{eq:1Dnonlin}
\end{align}
Similar to the example in  \cref{sec:1Dlin}, the measurement noise variance is $\sigma_\epsilon^2=0.05^2$, the measurement points are selected by maximin designs, and the test set $\{\bm{x}_i\}_{i=1}^{n_{\rm test}}$ contains $n_{\rm test}=500$ equally-spaced points for computing RMSE and MIS.

\begin{figure}
\centering
\includegraphics[width=0.99\textwidth]{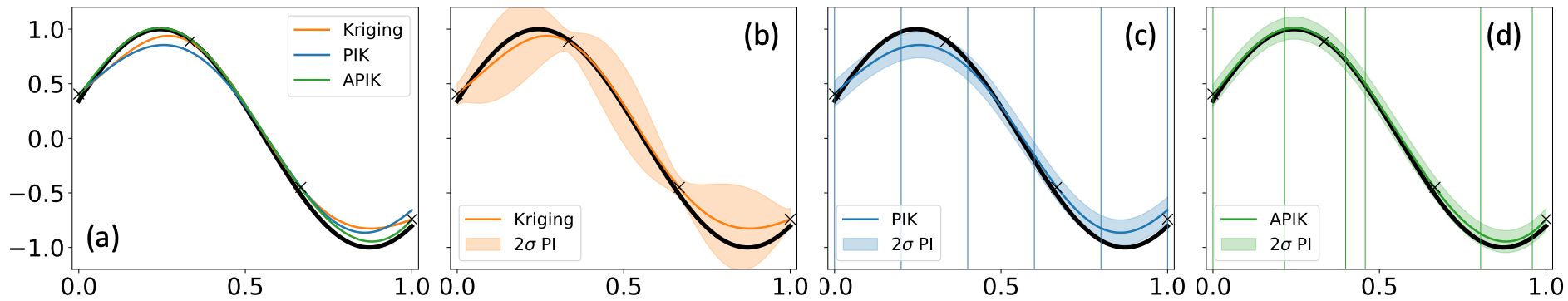}
\caption{\label{fig:NL1D} A 1D example with the nonlinear PDE \eqref{eq:1Dnonlin}: A comparison of the standard kriging method (b), PIK method (c), and the APIK method (d). The solid lines mark the predictive means, the vertical lines mark the locations of PDE points, and the shades outline the quantification of predictive uncertainty.}
\end{figure}

We consider the same three kriging-based methods --  standard kriging, PIK, and APIK.
As shown in \cref{fig:NL1D} (b), standard kriging performs reasonable; however, the prediction does not factorize in the PDE information at hand. The PIK model in \cref{fig:NL1D} (c), though incorporating the PDE information via equally-spaced PDE points, still struggles in providing improvements. This is because the PDE points in PIK are not specifically selected according to the measurement points. 

As shown in \cref{fig:NL1D} (d), APIK achieves noticeable improvements in accuracy and PI. The APIK method incorporates PDE information via a design procedure that considers the correlation between PDE points and measurement points. Therefore, the selected PDE points (vertical lines in \cref{fig:NL1D} (d)) both explore the whole space and exploit the regions where the PDE information is useful based on the measurement points. For example, the region  $[0.4,0.45]$ appears to be important to provide gradient information even with a measurement point nearby, whereas the region $[0.5,0.75]$ may not require additional gradient information since it is mostly linear.
By smartly leveraging PDE information, our APIK model achieves  (RMSE, MIS) of (0.041, 0.199), which is much smaller than (0.090, 0.511) for standard kriging and (0.102, 0.600) for PIK.

\cref{tab:RMSE1DNonLi} compares the predictive performance of the three considered methods, with measurement data sizes ranging from $n\in\{3,4,5,7,10\}$ and the corresponding PDE data sizes $m \in \{5,6,7,10,14\}$.  For predictive accuracy, APIK  achieves the smallest RMSEs with different data sizes. The improvements are particularly noticeable when the measurement sizes are small. For uncertainty quantification,  APIK  provides consistent improvements against both standard kriging and PIK. This is again because the proposed AIPK method incorporates and properly leverages the PDE information via carefully designed PDE points, and thereby demonstrating better learning performances.

\begin{figure}
\centering
\includegraphics[width=0.99\textwidth]{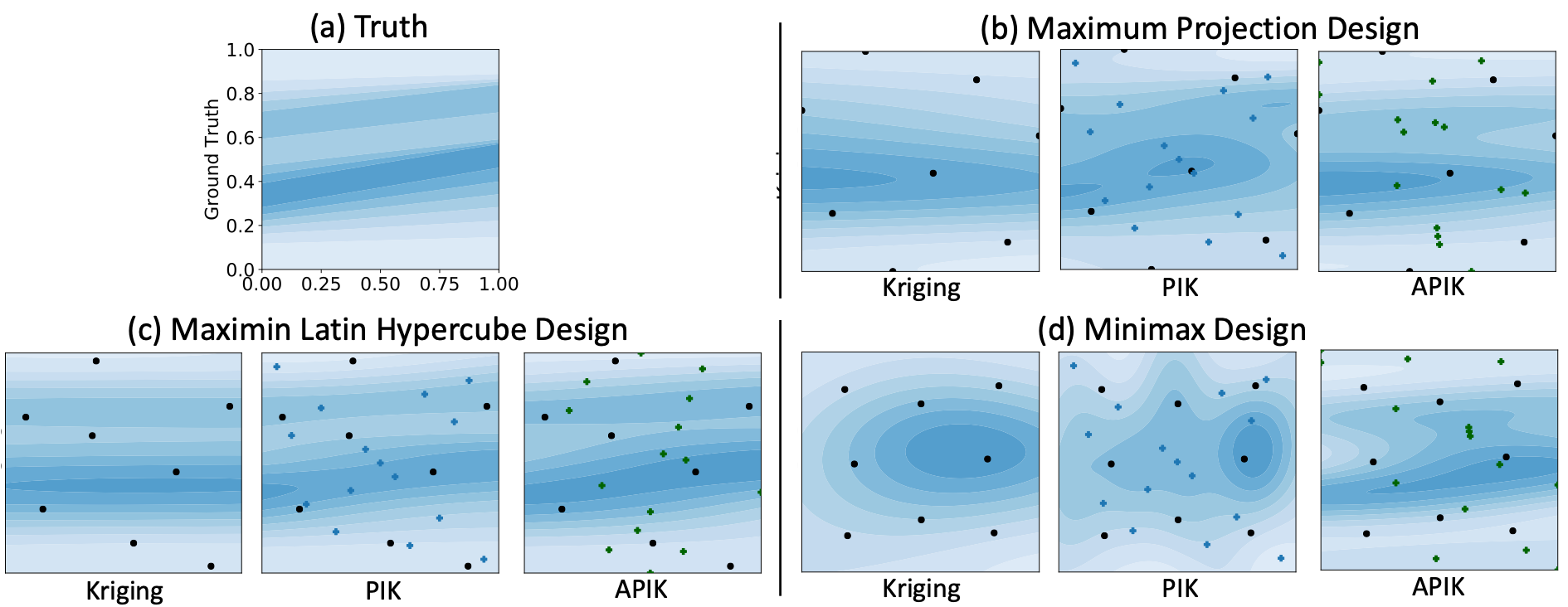}
\caption{\label{fig:IB} Shack wave case study: (a) shows the ground truth velocity field; 
(b), (c), and (d) show the predicted velocity fields by the three considered methods, with maximum projection design, maximin Latin hypercube design, and minimax design, respectively. The black dots mark the measurement prints, and the blue crosses mark the PDE prints.}
\end{figure}

\subsection{Shock wave case study} \label{sec:InBEquation}
We consider the nonlinear  inviscid Burger's equation \cite{burgers1948mathematical}:
\begin{align}
    \frac{\partial y(\bm{x}) }{\partial t} + y(\bm{x}) \frac{\partial y (\bm{x})}{\partial z}=0,\quad \bm{x} = [t,z]\in[0,1]^2.
    \label{eq:IBEq}
\end{align}
Here, $y(\bm{x})$ is the velocity field under consideration, with the input vector $\bm{x}$ containing the temporal and spatial coordinates. Burger's equation is a fundamental PDE for conservation equations that can develop discontinuities ({\it i.e.}, shock waves), occurring in various areas of applied mathematics, such as fluid mechanics, nonlinear acoustics, and traffic flow.

\begin{table}[!t]
\scriptsize
\centering
\begin{tabular}{c  | c c c| c c c} 
\toprule
  & \multicolumn{3}{c|}{RMSE}   &  \multicolumn{3}{c}{MIS}\\
Design & Kriging & PIK & APIK & Kriging &PIK & APIK\\
\midrule
 MaxPro &  0.0303  & 0.0233  & \textbf{0.0195} & 0.3068  &  0.2310 &  \textbf{0.1131} \\
 MmLHS & 0.0251		  & 0.0229 & \textbf{0.0193} &  0.3312  &  	0.2046 &  \textbf{0.1266} \\
 mM & 0.0389  & 0.0365  & \textbf{0.0266} & 0.4185  &  0.3514 &  \textbf{0.2051} \\
\toprule
\end{tabular}
\caption{A comparison of the predictive performance (i.e., RMSE and MIS) under the three considered methods in the shock wave case study.}
\label{tab:MIS1DNonLi}
\end{table}

As exampled in \cref{sec:nonlinearPDE}, the inviscid Burger's equation \eqref{eq:IBEq} is a nonlinear PDE due to an interaction term $y(\bm{x})\frac{\partial y(\bm{x})}{\partial z}$. 
The goal is to understand the velocity field $y(\bm{x})$ over the whole 2D spatial and temporal domain $[t,z]\in[0,1]^2$, with $n=8$  measurement points and $m=15$ PDE points.
The actual sensor measurement locations are selected via three popular space-filling designs --  the maximal projection design \cite{joseph2015maximum}, the minimax design using clustering \cite{mak2018minimax}, and the maximin Latin hypercube design \cite{morris1995exploratory},
thanks to the implementations in the corresponding \textsf{R} packages.

The detailed setup is as follows. We first assume (\textit{unknown} in reality, see \cref{sec:solvePDE}) an initial condition $y([t=0,z])$, and obtain the whole field $y(\bm{x})$ by solving inviscid Burger's equation \eqref{eq:IBEq} via  numerical finite difference method \cite{strikwerda2004finite}. The obtained velocity field $y(\bm{x})$ will be the ground truth function. The test set contains $201\times 201$ grid points over the input space $[0,1]^2$. We then simulate $n=8$ measurement data with $y_i = y(\bm{x}_i)+\epsilon_i, i=1,\cdots,8$, with additional measurement errors i.i.d. normally distributed $\epsilon_i\sim \mathcal{N}(0,1)$. The simulated measurement data $\{y_i\}_{i=1}^n$ are then used to estimate the standard kriging model. For PIK, we set the PDE points from an $m=15$-run Sobol' sequence since it provides a simple way of constructing a space-filling design.
The PIK model is estimated via an EM algorithm, similar to \texttt{LearnAPIK} ({\it i.e.}, \cref{alg:nonlinearAPIK}) yet without the design step. 
Finally, we estimate the APIK model via \texttt{LearnAPIK}. 

\cref{fig:IB} shows the predictive results with the three measurement data designs, using the three considered methods. \cref{tab:MIS1DNonLi} lists the corresponding RMSEs and MISs. 
Compared to standard kriging,  PIK slightly improves the predictive accuracy thanks to the additional PDE information. In addition, APIK further improves the learning performance with at least $20\%$ smaller RMSEs and $60\%$ smaller MISs, compare to both baseline methods.
This is again because our APIK method designs a PDE data set that minimizes the predictive uncertainty. For all three designs shown in \cref{fig:IB} (b-d), the PDE points selected by APIK not only explore the whole input space but also exploit the locations where the PDE information is the most important. For example, the velocity change in the vertical coordinate is larger than that in the horizontal coordinate in this application. The APIK method accordingly samples clusters of PDE points covering a wide range of vertical coordinates yet with similar horizontal coordinates. By properly leveraging the PDE information via the PDE points, APIK achieves noticeable improvements compared to standard kriging and PIK.

\begin{figure}
\centering
\includegraphics[width=0.45\textwidth]{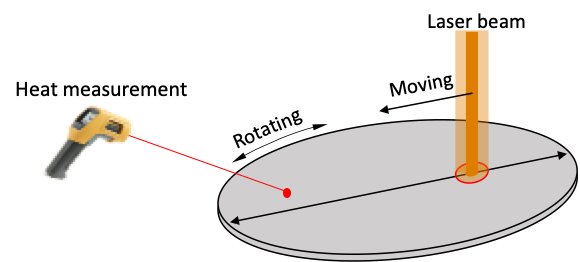}
\caption{\label{fig:WaferIll} An illustration of the laser heating process, and the assassinate temperature measurement in the wafer manufacturing case study.}
\end{figure}

\subsection{Laser heating case study} \label{sec:HeatTrans} Finally, we look into the laser heating process, which is an important step in the semiconductor wafer manufacturing process \cite{chen2019adaptive,chen2019hierarchical}. Wafer manufacturing involves processing silicon wafers in a series of refinement stages to be used as circuit chips.
Among these stages, thermal processing is one of the most important stages
since it facilitates the necessary chemical reactions and allows for surface oxidation. 
\cref{fig:WaferIll} illustrates the typical thermal processing procedure: A laser beam (in orange) is moved back and forth over a rotating wafer. Here, industrial engineers wish to understand the temperature field of the wafer over the whole heating process, which would provide a better understanding of possible thermal stresses and thereby the quality of final circuit chips.

During the heating process, the engineers can conduct temperature measurements using the heat measurement gun (also see \cref{fig:WaferIll}). However, since the heating process is typically short, only $n=12$ measurement data are collected. Those measurement locations are selected by a minimax design for its flexibility in the design space shape and good predictive performance. Furthermore, the heat transfer process is governed by the Fourier equation:
\begin{align}
     \frac{\partial y }{\partial t}- \beta \frac{\partial^2 y }{\partial z_1^2} - \beta \frac{\partial^2 y }{\partial z_2^2} = 0 .
    \label{eq:FE}
\end{align}
According to the discussion in \cref{sec:APIKmsize}, we select $m=25$ PDE points. The detailed procedure is similar to that in \cref{sec:InBEquation}. The only difference is that we use COMSOL multiphysics \cite{dickinson2014comsol}, a finite element analysis software, to obtain the temperature profile on the wafer, with no heat flux boundary condition assumed.

\begin{figure}
\centering
\includegraphics[width=0.99\textwidth]{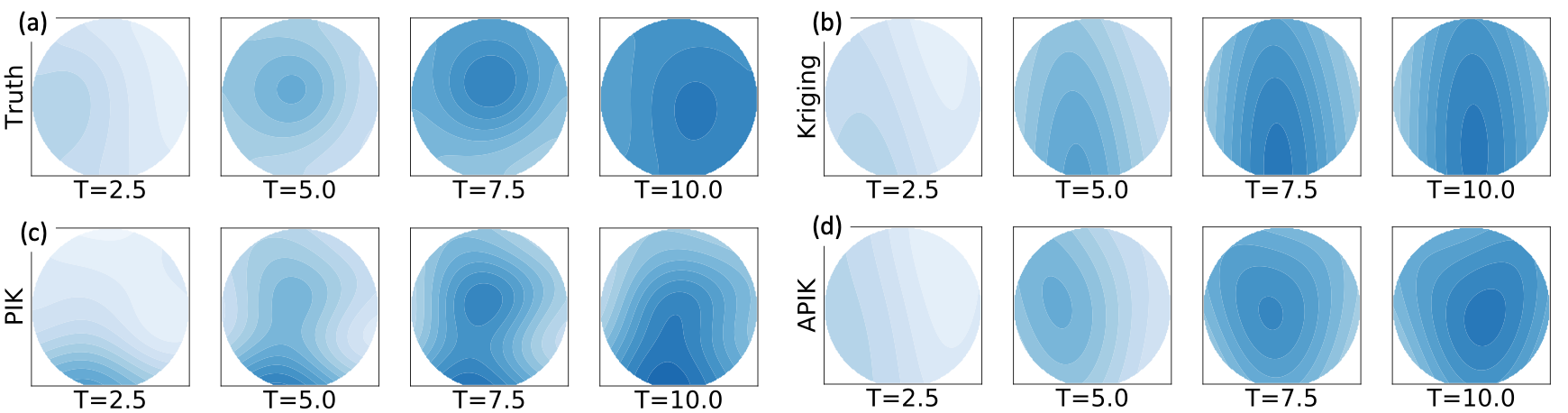}
\caption{\label{fig:Wafer} A comparison of the temperature contours at different time frames for the wafer manufacturing case study: (a) ground truth, (b) prediction by standard kriging, (c) prediction by PIK, and (d) prediction by APIK.}
\end{figure}

\cref{fig:Wafer} shows the four time frames of the ground truth wafer temperature profile during the heating process, and the predicted profiles via the three considered methods. While PIK already shows improvements against standard kriging by incorporating the PDE information, the proposed APIK method further improves the learning performance by exploring the correlation between measurement points and PDE points. 
Adopting a set of PDE points that smartly leverage PDE information, our APIK method achieves (RMSE, MIS) of (0.0308, 0.1173), which is much smaller than (0.0397, 0.1513) for standard kriging and (0.0333, 0.1376) for PIK. 
Furthermore, as shown in \cref{fig:Wafer},  APIK appears to better capture the divergence field of the temperature, which is known to be critical in introducing the thermal stress and affecting wafer quality \cite{eslami2013theory}.

\section{Conclusion}
\label{sec:summary}

In this work, we propose a new physics-informed learning method that combines partial differential equations (PDEs) and measurement data. We adopt the popular \textit{kriging} method, {\it i.e.}, Gaussian process regression, for its flexibility and closed-form expressions for both predictive mean and  uncertainty quantification. We first propose a PDE Informed Kriging model (PIK) to incorporate both linear and nonlinear PDEs via a pre-selected set of PDE points. Specifically, for nonlinear PDEs, we propose introducing latent variables to linearize the PDEs, and therefore, PIK enjoys the elegant posterior prediction framework similar to standard kriging. We then develop an APIK framework to also actively design the PDE points. The selected PDE points not only explore the whole input domain but also exploit locations where the PDE information is complementary to the measurement data. To estimate the APIK model, we present an expectation-maximization algorithm with fast updates, and also provide a heuristic method for the size of PDE points.
Finally, we demonstrate the effectiveness of the APIK method in two synthetic examples and two real-world case studies.

\appendix

\section{Proof for \cref{prop:PIKlinear}}
 \label{app:PIKlinear}
 
We derive here the prediction formula for PIK with linear PDEs. Following \cref{le:JGP}, we have the joint distribution for measurement data $\bm{y}_{1:n}$, outputs of the PDE process $\bm{y}^\mathcal{F}_{1:m}$, and the prediction $y(\bm{x}_{\rm new})$ at a new location $\bm{x}_{\rm new}$
\begin{align}
    \begin{bmatrix}
    \bm{y}_{1:n}  \vspace{0.1cm} \\
    \bm{y}^\mathcal{F}_{1:m}  \vspace{0.1cm}\\
    y(\bm{x}_{\rm new})
\end{bmatrix} \sim 
\mathcal{N}\left(\begin{bmatrix}
    \bm{P}  \vspace{0.1cm}\\
    \bm{P}_\mathcal{F}  \vspace{0.1cm}\\
    \bm{p}(\bm{x}_{\rm new})
\end{bmatrix}\boldsymbol\beta , \; \sigma^2 \begin{bmatrix}
    \bm{R}_{\mathcal{I}\mathcal{I}} & \bm{R}_{\mathcal{I}\mathcal{F}} &
    \bm{r}(\bm{x}_{\rm new})
     \vspace{0.1cm} \\
    \bm{R}_{\mathcal{I}\mathcal{F}}^\top  & \bm{R}_{\mathcal{F}\mathcal{F}} &
    \bm{r}_{\mathcal{F}} (\bm{x}_{\rm new})
    \vspace{0.1cm}  \\
     \bm{r} ^\top(\bm{x}_{\rm new})  & \bm{r}_{\mathcal{F}} ^\top(\bm{x}_{\rm new}) &
    1
\end{bmatrix}
\right),
\end{align}
Here, $\bm{r}_{\mathcal{F}} (\bm{x}) = [R_{\boldsymbol\theta}^{\mathcal{I}\mathcal{F}}(\bm{x},\bm{x}^\mathcal{F}_1),\cdots,R_{\boldsymbol\theta}^{\mathcal{I}\mathcal{F}}(\bm{x},\bm{x}^\mathcal{F}_m)]^\top$ is the correlation vector, matrices $\bm{R}_{\mathcal{I}\mathcal{I}}$, $\bm{R}_{\mathcal{I}\mathcal{F}}$, and $\bm{R}_{\mathcal{F}\mathcal{F}}$ are defined in \cref{le:JGP}. Using the conditional Gaussian formulation, we have 
\begin{align}
\hat{y}(\mathbf{x}_{\rm new}) &:=\mathbb{E} \left[y(\bm{x}_{\rm new})|\tilde{\bm{y}}_{n+m}\right]  =  \bm{p}^\top(\bm{x}_{\rm new})\boldsymbol\beta + \boldsymbol\gamma^\top(\bm{x}_{\rm new}) \boldsymbol\Gamma_{\boldsymbol\theta}^{-1}(\tilde{\bm{y}}_{n+m}-\tilde{\bm{P}}\boldsymbol\beta),\\
s^2(\mathbf{x}_{\rm new}) &:= \text{\rm Var} \left[y(\bm{x}_{\rm new})|\tilde{\bm{y}}_{n+m}\right]  =  \sigma^2 (1- \boldsymbol\gamma^\top(\bm{x}_{\rm new})\boldsymbol\Gamma_{\boldsymbol\theta}^{-1}\boldsymbol\gamma(\bm{x}_{\rm new})),
\end{align}
where $\tilde{\bm{P}}= [\bm{P}^\top,\bm{P}_\mathcal{F}^\top]^\top$ is the model matrix, and $\boldsymbol\gamma(\bm{x}) = [\bm{r}^\top(\bm{x}),\bm{r}_{\mathcal{F}}^\top (\bm{x})]^\top$ is the correlation vector, and $\boldsymbol\Gamma_{\boldsymbol\theta}$ is the corresponding correlation matrix.

\section{Proof for \cref{cor:linearPDE}}
 \label{app:linearPDECor}
The goal is to find the predictive distribution of $y^{\mathcal{G}}(\cdot) := \mathcal{G}_{\bm{x}}[y](\cdot)$.
Similarly to \cref{app:PIKlinear}, we have the joint distribution of the same three parts \begin{align}
    \begin{bmatrix}
    \bm{y}_{1:n}   \vspace{0.1cm}\\
    \bm{y}^\mathcal{F}_{1:m} \vspace{0.1cm}\\
    y^\mathcal{G}(\bm{x}_{\rm new})
\end{bmatrix} \sim 
\mathcal{N}\left(\begin{bmatrix}
    \bm{P} \vspace{0.1cm}\\
    \bm{P}_\mathcal{F} \vspace{0.1cm}\\
    \bm{p}_\mathcal{G}(\bm{x}_{\rm new})
\end{bmatrix}\boldsymbol\beta , \; \sigma^2 \begin{bmatrix}
    \bm{R}_{\mathcal{I}\mathcal{I}} & \bm{R}_{\mathcal{I}\mathcal{F}} &
    \bm{r}_{\mathcal{G}}(\bm{x}_{\rm new})
     \vspace{0.1cm}\\
    \bm{R}_{\mathcal{I}\mathcal{F}}^\top  & \bm{R}_{\mathcal{F}\mathcal{F}} &
    \bm{r}_{\mathcal{F}\mathcal{G}} (\bm{x}_{\rm new})
    \vspace{0.1cm} \\
     \bm{r}_{\mathcal{G}} ^\top(\bm{x}_{\rm new})  & \bm{r}_{\mathcal{F}\mathcal{G}} ^\top(\bm{x}_{\rm new}) &
    r_{\mathcal{G}}(\bm{x}_{\rm new}) 
\end{bmatrix}
\right),
\end{align}
Here, $ \bm{r}_{\mathcal{G}} (\bm{x}) = [R_{\boldsymbol\theta}^{\mathcal{I}\mathcal{G}}(\bm{x},\bm{x}_1),\cdots,R_{\boldsymbol\theta}^{\mathcal{I}\mathcal{G}}(\bm{x},\bm{x}_n)]^\top$ is the correlation vector for the  measurement data, and $ \bm{r}_{\mathcal{F}\mathcal{G}} (\bm{x}) = [R_{\boldsymbol\theta}^{\mathcal{F}\mathcal{G}}(\bm{x},\bm{x}^\mathcal{F}_1),\cdots,R_{\boldsymbol\theta}^{\mathcal{F}\mathcal{G}}(\bm{x},\bm{x}^\mathcal{F}_m)]^\top$ is the correlation vector for the  PDE data, with the notation for derivative correlation functions
\begin{align}
R_{\boldsymbol\theta}^{\mathcal{I}\mathcal{G}}(\bm{x},\bm{x}'): =  (\mathcal{I}_{\bm{x}} \times \mathcal{G}_{z'})[R_{\boldsymbol\theta}](\bm{x},\bm{x}'), \quad R_{\boldsymbol\theta}^{\mathcal{F}\mathcal{G}}(\bm{x},\bm{x}') := (\mathcal{F}_{\bm{x}} \times \mathcal{G}_{z'})[R_{\boldsymbol\theta}](\bm{x},\bm{x}')
\end{align}
following from \eqref{eq:DerivativeProcess}. Furthermore, $\sigma^2 r_{\mathcal{G}}(\bm{x})=  \sigma^2 R_{\boldsymbol\theta}^{\mathcal{G}\mathcal{G}}(\bm{x},\bm{x})$ is the prior process variance at $\bm{x}$.
The posterior distribution of the prediction $y^{\mathcal{G}}(\bm{x}_{\rm new})$ at a new location $\bm{x}_{\rm new}$ is Gaussian with following mean and variance
\begin{align}
\mathbb{E} \left[y^{\mathcal{G}}(\bm{x}_{\rm new})|\tilde{\bm{y}}_{n+m}\right] & =  \bm{p}_\mathcal{G}^\top(\bm{x}_{\rm new})\boldsymbol\beta + \boldsymbol\gamma_\mathcal{G}^\top(\bm{x}_{\rm new}) \boldsymbol\Gamma_{\boldsymbol\theta}^{-1}(\tilde{\bm{y}}_{n+m}-\tilde{\bm{P}}\boldsymbol\beta), \\
\text{\rm Var}\left[y^{\mathcal{G}}(\bm{x}_{\rm new})|\tilde{\bm{y}}_{n+m}\right] & =  \sigma^2 (r_{\mathcal{G}}(\bm{x})- \boldsymbol\gamma_\mathcal{G}^\top(\bm{x}_{\rm new})\boldsymbol\Gamma_{\boldsymbol\theta}^{-1}\boldsymbol\gamma_\mathcal{G}(\bm{x}_{\rm new})).
\end{align}
Here, coefficients $\bm{p}_\mathcal{G}(\bm{x})=[\mathcal{G}_{\bm{x}}[p_1](\bm{x}),\cdots, \mathcal{G}_{\bm{x}}[p_q](\bm{x})]^\top$ consists of the derivatives of the $q$ basis functions, and  $\boldsymbol\gamma_\mathcal{G}(\bm{x}) = [ \bm{r}_{\mathcal{G}}^\top,  \bm{r}_{\mathcal{F}\mathcal{G}}^\top]^\top$ is the correlation vector.

\section{Proof for \cref{prop:PIKnonlinear}}
 \label{app:PIKnonlinear}
Now, we derive the prediction formula for PIK with \textit{nonlinear} PDEs. As discussed in \cref{sec:nonlinearPDE}, we have linearize the \textit{nonlinear} differential operator $\mathcal{F}[\cdot]$ to a \textit{linear} differential operator $\bar{\mathcal{F}}[\cdot]$ via a set of derivatives $\partial$ and the corresponding latent variable $\bm{z}$. We have the following joint distribution for measurement data $\bm{y}_{1:n}$, outputs of the \textit{linearized} PDE process $\bm{y}^{\bar{\mathcal{F}}}_{1:m}$, the latent variable $\bm{z}$, and the prediction $y(\bm{x}_{\rm new})$ at a new location $\bm{x}_{\rm new}$
\begin{align}
    \begin{bmatrix}
    \bm{y}_{1:n}  \vspace{0.1cm} \\
    \bm{y}^{\bar{\mathcal{F}}}_{1:m} \vspace{0.1cm}\\
    \bm{z} \vspace{0.1cm}\\
    y(\bm{x}_{\rm new})
\end{bmatrix} \sim 
\mathcal{N}\left(\begin{bmatrix}
    \bm{P} \vspace{0.1cm}\\
    \bm{P}_{\bar{\mathcal{F}}} \vspace{0.1cm}\\
    \bm{P}_\partial \vspace{0.1cm}\\
    \bm{p}(\bm{x}_{\rm new})
\end{bmatrix}\boldsymbol\beta , \; \sigma^2 \begin{bmatrix}
    \bm{R}_{\mathcal{I}\mathcal{I}} & \bm{R}_{\mathcal{I}\bar{\mathcal{F}}} &
    \bm{R}_{\mathcal{I}\partial} &
    \bm{r}(\bm{x}_{\rm new})
    \vspace{0.1cm} \\
    \bm{R}_{\mathcal{I}\bar{\mathcal{F}}}^\top  & \bm{R}_{\bar{\mathcal{F}}\bar{\mathcal{F}}} &
    \bm{R}_{\bar{\mathcal{F}}\partial} &
    \bm{r}_{\bar{\mathcal{F}}} (\bm{x}_{\rm new})
    \vspace{0.1cm} \\
    \bm{R}_{\mathcal{I}\partial}^\top &
    \bm{R}_{\bar{\mathcal{F}}\partial}^\top&
    \bm{R}_{\partial\partial} &
     \bm{r}_{\partial} (\bm{x}_{\rm new})
   \vspace{0.1cm}  \\
     \bm{r} ^\top(\bm{x}_{\rm new})  & \bm{r}_{\bar{\mathcal{F}}}^\top (\bm{x}_{\rm new}) & 
     \bm{r}_{\partial}^\top (\bm{x}_{\rm new})&
    1
\end{bmatrix}
\right).
\end{align}
Here, $\bm{r}_{\partial} (\bm{x}) = [R_{\boldsymbol\theta}^{\mathcal{I}\partial}(\bm{x},\bm{x}^\mathcal{F}_1),\cdots,R_{\boldsymbol\theta}^{\mathcal{I}\partial}(\bm{x},\bm{x}^\mathcal{F}_m)]^\top$ is the correlation vector for the latent variable $\bm{z}$,  $\bm{r}_{\bar{\mathcal{F}}} (\bm{x}) = [R_{\boldsymbol\theta}^{\mathcal{I}\bar{\mathcal{F}}}(\bm{x},\bm{x}^\mathcal{F}_1),\cdots,R_{\boldsymbol\theta}^{\mathcal{I}\bar{\mathcal{F}}}(\bm{x},\bm{x}^\mathcal{F}_m)]^\top$ is the correlation vector for the outputs of the \textit{linearized} PDE process $\bar{\mathcal{F}}[\cdot]$, and $\bm{P}_\partial= [\bm{p}_\partial(\bm{x}^\mathcal{F}_1),\cdots, \bm{p}_\partial(\bm{x}^\mathcal{F}_m)]^\top$ is the model matrix, with $\bm{p}_\partial(\bm{x})=[\partial[p_1](\bm{x}),\cdots, \partial[p_q](\bm{x})]^\top$ containing the $q$ derivatives of the basis functions. Note that, for notation simplicity, we consider $K=|\partial|=1$; For the cases that $K>1$, $\bm{r}_{\partial} (\bm{x})$ and $\bm{P}_\partial$ would be the concatenation of all the derivatives in the set $\partial$. Furthermore, the correlation matrices take the following forms:
\begin{align}
    \bm{R}_{\mathcal{I}\bar{\mathcal{F}}} = {[R_{\boldsymbol\theta}^{\mathcal{I}\bar{\mathcal{F}}}(\bm{x}_i,\bm{x}_j^\mathcal{F})]_{i=1}^n}_{j=1}^m \quad & \text{with}  \quad R_{\boldsymbol\theta}^{\mathcal{I}\bar{\mathcal{F}}}(\bm{x},\bm{x}') := (\mathcal{I}_{\bm{x}} \times \bar{\mathcal{F}}_{\bm{x}'})[R_{\boldsymbol\theta}](\bm{x},\bm{x}'); \\
    \bm{R}_{\bar{\mathcal{F}}\bar{\mathcal{F}}} = {[R_{\boldsymbol\theta}^{\bar{\mathcal{F}}\bar{\mathcal{F}}}(\bm{x}_i,\bm{x}_j^\mathcal{F})]_{i=1}^n}_{j=1}^m \quad & \text{with}  \quad R_{\boldsymbol\theta}^{\bar{\mathcal{F}}\bar{\mathcal{F}}}(\bm{x},\bm{x}') := (\bar{\mathcal{F}} \times \bar{\mathcal{F}}_{\bm{x}'})[R_{\boldsymbol\theta}](\bm{x},\bm{x}');\\
    \bm{R}_{\mathcal{I}\partial} = {[R_{\boldsymbol\theta}^{\mathcal{I}\partial}(\bm{x}_i,\bm{x}_j^\mathcal{F})]_{i=1}^n}_{j=1}^m \quad & \text{with}  \quad R_{\boldsymbol\theta}^{\mathcal{I}\partial}(\bm{x},\bm{x}') := (\mathcal{I}_{\bm{x}} \times \partial_{z'})[R_{\boldsymbol\theta}](\bm{x},\bm{x}');\\
    \bm{R}_{\bar{\mathcal{F}}\partial} = {[R_{\boldsymbol\theta}^{\bar{\mathcal{F}}\partial}(\bm{x}_i,\bm{x}_j^\mathcal{F})]_{i=1}^n}_{j=1}^m \quad & \text{with}  \quad R_{\boldsymbol\theta}^{\bar{\mathcal{F}}\partial}(\bm{x},\bm{x}') := ( \bar{\mathcal{F}}_{\bm{x}}\times \partial_{z'})[R_{\boldsymbol\theta}](\bm{x},\bm{x}'); \\
    \bm{R}_{\partial\partial} = {[R_{\boldsymbol\theta}^{\partial\partial}(\bm{x}_i,\bm{x}_j^\mathcal{F})]_{i=1}^n}_{j=1}^m \quad & \text{with}  \quad R_{\boldsymbol\theta}^{\partial\partial}(\bm{x},\bm{x}') := (\partial_{z} \times \partial_{z'})[R_{\boldsymbol\theta}](\bm{x},\bm{x}').
\end{align}

Then the posterior mean of the prediction $y(\bm{x}_{\rm new})$ at a new input location $\bm{x}_{\rm new}$, conditional on both the measurement data $\bm{y}_{1:n}$ and the outputs of the linearized PDE process $\bm{y}^{\bar{\mathcal{F}}}_{1:m}$ can be evaluated by the law of total expectation: 
\begin{align}
\hat{y}(\mathbf{x}_{\rm new}) &=\mathbb{E} \left[y(\bm{x}_{\rm new})|\tilde{\bm{y}}_{n+m}\right]  =\mathbb{E}\left[\mathbb{E}  \left[y(\bm{x}_{\rm new})|\tilde{\bm{y}}_{n+m},\bm{Z
}\right]|\tilde{\bm{y}}_{n+m}\right] \label{eq:app:E} \\
&= \mathbb{E} \left[\bm{p}^\top(\bm{x}_{\rm new})\boldsymbol\beta + \boldsymbol\gamma^\top(\bm{x}_{\rm new}) \boldsymbol\Gamma_{\boldsymbol\theta}^{-1}([\tilde{\bm{y}}_{n+m}^\top,\bm{z}^\top]^\top-\tilde{\bm{P}}\boldsymbol\beta)|\tilde{\bm{y}}_{n+m}\right] \\
&=\bm{p}^\top(\bm{x}_{\rm new})\boldsymbol\beta + \boldsymbol\gamma^\top(\bm{x}_{\rm new}) \boldsymbol\Gamma_{\boldsymbol\theta}^{-1}([\tilde{\bm{y}}^\top_{n+m},\mathbb{E}[\bm{z}|\tilde{
\bm{y}}_{n+m}]^\top]^\top-\tilde{\bm{P}}\boldsymbol\beta).
\end{align}

Similarly, the posterior variance of the prediction $y(\bm{x}_{\rm new})$ at a new input location $\bm{x}_{\rm new}$, conditional on $\bm{y}_{1:n}$ and $\bm{y}^{\bar{\mathcal{F}}}_{1:m}$ is 
\begin{align}
s^2(\mathbf{x}_{\rm new}) &= \text{Var} \left[y(\bm{x}_{\rm new})|\tilde{\bm{y}}_{n+m}\right]\\
&= \mathbb{E} \left[\text{Var} \left[y(\bm{x}_{\rm new})|\tilde{\bm{y}}_{n+m},\bm{z}\right] |\tilde{\bm{y}}_{n+m}\right]+ \text{Var}\left[\mathbb{E} \left[y(\bm{x}_{\rm new})|\tilde{\bm{y}}_{n+m},\bm{z}\right] |\tilde{\bm{y}}_{n+m}\right] \label{eq:app:var}\\
&= \text{Var} \left[y(\bm{x}_{\rm new})|\tilde{\bm{y}}_{n+m},\bm{z}\right]+ \text{Var}\left[\mathbb{E} \left[y(\bm{x}_{\rm new})|\tilde{\bm{y}}_{n+m},\bm{z}\right] |\tilde{\bm{y}}_{n+m}\right]\\
&=  \sigma^2- \boldsymbol\gamma^\top(\bm{x}_{\rm new})\boldsymbol\Gamma_{\boldsymbol\theta}^{-1}\left[\sigma^2\boldsymbol\gamma(\bm{x}_{\rm new})-\text{diag}(\bm{0}_{n+m},\text{Var}[\bm{z}|\tilde{\bm{y}}_{n+m}]) \boldsymbol\Gamma_{\boldsymbol\theta}^{-1}\boldsymbol\gamma(\bm{x}_{\rm new})\right].
\end{align}
Note that the law of total variance is used in \eqref{eq:app:var}.

\section{Likelihood function}
\label{app:like}
Without loss of generality, we set the size of measurement data $n=1$, {\it i.e.}, $\bm{y}_{1:n}=y_1$, and the size of the PDE data  $m=1$, {\it i.e.}, $\bm{y}^\mathcal{F}_{1:m}=w_1$.   Denote the random variable for the measurement observation $Y$, that for the latent variable $Z$, that for observation with the \textit{nonlinear} PDE $Y^\mathcal{F}$ and that for the corresponding \textit{linearized} PDE $\bar{Y}$.
According to the definition of likelihood, we have 
\begin{align}
     l(\boldsymbol\phi;\tilde{\bm{y}}_{n+m},\bm{z}) 
     =  -\log\{\mathbb{P}[Y=y_1,Y^\mathcal{F}=w_1, Z=z|\boldsymbol\phi]\}
\end{align}
Now introduce the linearized PDE with differential operator $\bar{\mathcal{F}}$ and the associated output random variable $\bar{Y}$:
\begin{align}
      \mathbb{P}[Y=y_1,Y^\mathcal{F}=w_1, Z=z|\boldsymbol\phi] =  \mathbb{P}[Y=y_1,\bar{Y}=w_1, Z=z|\boldsymbol\phi]
 \end{align}
 Note that thanks to the linearization, the joint probability density  function of $[Y,\bar{Y}, Z]$ is in Gaussian form with mean $\tilde{\bm{P}}\boldsymbol\beta$ and the correlation $\boldsymbol\Gamma_{\boldsymbol\theta}$. Note that the  correlation $\boldsymbol\Gamma_{\boldsymbol\theta}$ is a function of the latent value $z$. Finally, we have the negative log-likelihood $     l(\boldsymbol\phi;\tilde{\bm{y}}_{n+m},\bm{z})$
  \begin{align}
     \log( \det(\sigma^2\boldsymbol\Gamma_{\boldsymbol\theta})) +\frac{1}{\sigma^2}\left(\left[\tilde{\bm{y}}^\top_{n+m},
\bm{z}^\top\right]^\top-\tilde{\bm{P}}\boldsymbol\beta\right)^\top\boldsymbol\Gamma_{\boldsymbol\theta}^{-1}\left(\left[\tilde{\bm{y}}^\top_{n+m},
\bm{z}^\top\right]^\top-\tilde{\bm{P}}\boldsymbol\beta\right).
\end{align}

\section{Proof for \cref{prop:IMSEPDE}}
 \label{app:IMSEPDE} 
Since the correlation function $R_{\boldsymbol \theta}$ is sufficiently smooth and $R^{\mathcal{F}\mathcal{F}}_{\boldsymbol \theta}(\bm{x}, \bm{x}')$ exists, by Theorem 2.2.2 in \cite{adler1981geometry} and induction, we obtain that $y^{\mathcal{F}}$ also follows a Gaussian process with correlation function $R^{\mathcal{F}\mathcal{F}}_{\boldsymbol \theta}(\bm{x}, \bm{x}')$.

Then, by Mercer's expansion, we have $$R_{\boldsymbol\theta}^{\mathcal{F}\mathcal{F}}(\bm{x},\bm{x}') =\sum_{i=1}^\infty \xi_i \varepsilon_i(\bm{x})\varepsilon_i(\bm{x}'),$$
where $e_i(x)'$s are the orthonormal basis, {\it i.e.,} $\int_{\mathcal{X}} \varepsilon_i(\bm{x})\varepsilon_j(\bm{x}) d\bm{x} = 1$ if $i=j$, otherwise $0$. 

Furthermore, by \cref{le:IMSEMeansure}, we obtain the desired results.

\section{Proof for \cref{th:fastupdate}}

\label{app:fastupdate}
The correlation matrix $\boldsymbol\Gamma_{\boldsymbol\theta}$ can be written in 2-by-2 block matrix form (with switches in rows and columns)
\begin{align}
    \boldsymbol\Gamma_{\boldsymbol\theta} = 
    \left[
\begin{array}{cc}
\boldsymbol\Gamma_{\boldsymbol\theta-j}  & \boldsymbol\gamma_{-j} \vspace{0.15cm} \\
\boldsymbol\gamma_{-j}^\top  & d_j
\end{array}
\right].
\end{align}
Here, $\boldsymbol\Gamma_{\boldsymbol\theta-j}$ is the correlation matrix $\boldsymbol\Gamma_{\boldsymbol\theta}$ without the $j$-th row and column,  $\boldsymbol\gamma_{-j}(\bm{x})$ is the corresponding correlation vector with $n$ observed points and $(m-1)$ PDE points except for the $j$-th PDE point from \eqref{eq:APIK_E}, and $d_j=R_{\boldsymbol\theta}^{\mathcal{F}\mathcal{F}}(\bm{x}_j^{\mathcal{F}})$ is the value of the derivative of the correlation function at $\bm{x}_j^{\mathcal{F}}$. Adopting the formulation for the block matrix inversion, we have 
\begin{align}
    \left[
\begin{array}{cc}
\boldsymbol\Gamma_{\boldsymbol\theta-j}  & \boldsymbol\gamma_{-j} \vspace{0.15cm} \\ 
\boldsymbol\gamma_{-j}^\top  & d_j
\end{array}
\right]^{-1}
=   \left[
\begin{array}{cc}
\boldsymbol\Gamma_{\boldsymbol\theta-j}^{-1}+ \boldsymbol\Gamma_{\boldsymbol\theta-j}^{-1}\boldsymbol\gamma_{-j} d^-_j  \boldsymbol\gamma_{-j}^\top \boldsymbol\Gamma_{\boldsymbol\theta-j}^{-1} &- \boldsymbol\Gamma_{\boldsymbol\theta-j}^{-1}\boldsymbol\gamma_{-j}d^-_j   \vspace{0.15cm}\\
- \boldsymbol\gamma_{-j}^\top\boldsymbol\Gamma_{\boldsymbol\theta-j}^{-1}d^-_j  & d^-_j
\end{array}
\right].
\end{align}
where scalar $d^-_j = 1/(d_j-\boldsymbol\gamma_{-j}^\top\boldsymbol\Gamma_{\boldsymbol\theta-j}^{-1}\boldsymbol\gamma_{-j})$. Note that $d^-_j = 1/s^2_j(\bm{x}_j^{\mathcal{F}})$ holds according to the definition of the posterior variance in \eqref{eq:GPPDE_V}. Denoting vector $\bm{g}_j(\bm{x}) = \boldsymbol\Gamma_{\boldsymbol\theta-j}^{-1}\boldsymbol\gamma_{-j}(\bm{x})/s^2_j(\bm{x})$, the inverse $\boldsymbol\Gamma_{\boldsymbol\theta}^{-1}$ can be efficiently computed as follows:
\begin{align}
\boldsymbol\Gamma_{\boldsymbol\theta}^{-1} = 
\left[
\begin{array}{cc}
\boldsymbol\Gamma_{\boldsymbol\theta-j}^{-1} + \bm{g}_j(\bm{x}^{\mathcal{F}}_j)  \bm{g}_j(\bm{x}^{\mathcal{F}}_j)^\top s^2_j(\bm{x}^{\mathcal{F}}_j)   & \bm{g}_j(\bm{x}^{\mathcal{F}}_j)  \vspace{0.15cm} \\
 \bm{g}_j(\bm{x}^{\mathcal{F}}_j) ^\top    & 1/s^{2}_j(\bm{x}^{\mathcal{F}}_j) 
\end{array}
\right].
\end{align}

Note that the computation of $\boldsymbol\Gamma_{\boldsymbol\theta-j}^{-1}$ can also be obtained in a similar manner from $\boldsymbol\Gamma_{\boldsymbol\theta}^{-1}$. We leave this part out due to more cumbersome notation.


\bibliographystyle{siamplain}

\vfill
\end{document}